\documentclass[amsmath,aps,prd,amssymb, tightenlines, nofootinbib ]{revtex4-2}

\usepackage[utf8]{inputenc}
\usepackage{float}
\usepackage{graphics,graphicx}
\usepackage{dcolumn}
\usepackage{bm}
\usepackage{amsmath}
\usepackage{tensor}
\usepackage{slashed}
\usepackage{amsmath,amssymb,amsfonts,mathrsfs,bbold}
\usepackage{yfonts}
\usepackage{xcolor}
\usepackage{mathtools}
\usepackage{hyperref} 
\usepackage[framemethod=TikZ]{mdframed}
\usepackage{color}

\usepackage{multirow}
\usepackage{mathbbol}
\usepackage{mathtools}

\usepackage{slashed}
\usepackage{simpler-wick}

\usepackage[normalem]{ulem}

\usepackage{stackrel}
\usepackage[most]{tcolorbox}

\def\eq#1{{Eq.~(\ref{#1})}}
\def\eqs#1{{Eqs.~(\ref{#1})}}
\def\fig#1{{Fig.~\ref{#1}}}

\newcommand{\as}{\alpha_s}
\newcommand{\soz}{s_{10}}
\newcommand{\sto}{s_{21}}
\newcommand{\stt}{s_{32}}
\newcommand{\xoz}{x_{10}}
\newcommand{\xto}{x_{21}}
\newcommand{\xtt}{x_{32}}
\newcommand{\wint}{\int \frac{\mathrm{d}\omega}{2\pi i}}
\newcommand{\gint}{\int \frac{\mathrm{d}\gamma}{2\pi i}}

\newcommand{\dw}{\delta_{\omega}}
\newcommand{\gtwg}{G_{2\omega\gamma}}

\newcommand{\pd}{\partial}
\newcommand{\bas}{{\bar\alpha}_s}
\newcommand{\un}[1]{\underline{#1}}

\newcommand{\stf}{\frac{-\frac{3}{2}\omega\gamma + 4}{\gamma^2-\omega\gamma+1}}

\begin{document}

\title{Analytic Solution for the Revised Helicity Evolution at Small $x$ and Large $N_c\,$: \\ 
New Resummed Gluon-Gluon Polarized Anomalous Dimension and Intercept}

\author{Jeremy Borden}
         \email[Email: ]{borden.75@buckeyemail.osu.edu}
	\affiliation{Department of Physics, The Ohio State University, Columbus, OH 43210, USA}

\author{Yuri V. Kovchegov} 
         \email[Email: ]{kovchegov.1@osu.edu}
         \affiliation{Department of Physics, The Ohio State
           University, Columbus, OH 43210, USA}

\begin{abstract}
We construct an exact analytic solution of the revised small-$x$ helicity evolution equations derived in \cite{Cougoulic:2022gbk} based on the earlier work \cite{Kovchegov:2015pbl, Kovchegov:2018znm}. The equations we solve are obtained in the large-$N_c$ limit (with $N_c$ the number of quark colors) and are double-logarithmic (summing powers of $\alpha_s \ln^2(1/x)$ with $\as$ the strong coupling constant and $x$ the Bjorken $x$ variable). Our solution provides small-$x$, large-$N_c$ expressions for the flavor-singlet quark and gluon helicity parton distribution functions (PDFs) and for the $g_1$ structure function, with their leading small-$x$ asymptotics given by 
\begin{align}
        \Delta \Sigma (x, Q^2) \sim \Delta G (x, Q^2) 
    \sim g_1 (x, Q^2) \sim \left( \frac{1}{x} \right)^{\alpha_h} ,  \notag
\end{align}
where the exact analytic expression we obtain for the intercept $\alpha_h$ can be approximated by $\alpha_h = 3.66074 \, \sqrt{\frac{\alpha_s \, N_c}{2 \pi}}$. Our solution also yields an all-order (in $\alpha_s$) resummed small-$x$ anomalous dimension $\Delta \gamma_{GG} (\omega)$ which agrees with all the existing fixed-order calculations (to three loops). Notably, our anomalous dimension is different from that obtained in the infrared evolution equation framework developed earlier by Bartels, Ermolaev, and Ryskin (BER) \cite{Bartels:1996wc}, with the disagreement starting at four loops.  Despite the previously reported agreement at two decimal points based on the numerical solution of the same equations \cite{Cougoulic:2022gbk}, the intercept of our large-$N_c$ helicity evolution and that of BER disagree beyond that precision, with the BER intercept at large $N_c$ given by a different analytic expression from ours with the numerical value of $\alpha_h^{BER} = 3.66394 \, \sqrt{\frac{\alpha_s \, N_c}{2 \pi}}$. We speculate on the origin of this disagreement. 
\end{abstract}

\maketitle
\tableofcontents


\section{Introduction}
\label{sec:intro}

The proton spin puzzle \cite{Aidala:2012mv,Accardi:2012qut,Leader:2013jra,Aschenauer:2013woa,Aschenauer:2015eha,Boer:2011fh,Proceedings:2020eah,Ji:2020ena,AbdulKhalek:2021gbh} remains one of the fundamental open questions in our understanding of the hadronic structure.  The main question of the puzzle is 
how the spin of the proton is distributed among its quarks and gluons. This is best described by the spin sum rules, due to Jaffe and Manohar \cite{Jaffe:1989jz} and due to Ji \cite{Ji:1996ek}. The former reads 
\begin{equation}
S_q+L_q+S_G+L_G=\frac{1}{2},
\label{eqn:JM}
\end{equation}
where $S_q$ and $S_G$ are the contributions to the spin of the proton coming from the quark and gluon helicities, respectively, and $L_q$ and $L_G$ are the contributions due to the quark and gluon orbital angular momenta (OAM). 

The $S_q$ and $S_G$ can be written as integrals over the Bjorken $x$ variable, 
\begin{align}
S_q(Q^2) = \frac{1}{2} \int\limits_0^1 dx \; \Delta\Sigma(x,Q^2), \ \ \  \ \
S_G(Q^2) = \int\limits_0^1 dx \; \Delta G(x,Q^2),
\label{eqn:SqSG}
\end{align}
where the flavor-singlet helicity distribution is
\begin{equation}
\Delta\Sigma(x,Q^2) = \sum_{f=u,d,s,\ldots} 
\left[\Delta q_f (x,Q^2) + \Delta\bar{q}_f (x,Q^2)\right].
\label{eqn:DeltaSigma}
\end{equation}
Here $\Delta q_f (x,Q^2)$, $\Delta\bar{q}_f (x,Q^2)$ are the quark and anti-quark helicity distributions, respectively, while $\Delta G(x,Q^2)$ is the gluon helicity distribution function. The reader is referred to \cite{Accardi:2012qut,Leader:2013jra,Aschenauer:2013woa,Aschenauer:2015eha,Proceedings:2020eah,Ji:2020ena} for detailed reviews of the proton spin puzzle.  

One of the least-explored regions of phase space which may potentially contribute to $S_q(Q^2)$ and $S_G(Q^2)$ is at small $x$, in part due to the limited amount of available relevant data in that region and in part due to finite acceptance in $x$ of any given experiment, not allowing the exploration of helicity parton distribution functions (hPDFs) down to $x=0$, as required by Eqs.~\eqref{eqn:SqSG}. Thus, understanding and quantifying the amount of the proton's spin carried by small-$x$ partons is an integral part of the proton spin puzzle. Theoretical input appears to be necessary here: even future experiments, such as those to be performed at the Electron-Ion Collider (EIC) \cite{Accardi:2012qut,Boer:2011fh,Proceedings:2020eah,AbdulKhalek:2021gbh}, would only be able to probe the $x$-region down to some $x_{min}$, below which a theoretical extrapolation to lower $x$ would still be required to constrain the net amount of proton spin at small $x$.
 
In the perturbative quantum chromodynamics (pQCD) framework the first attempt at calculating the hPDFs at small $x$ was done by Bartels, Ermolaev and Ryskin (BER) \cite{Bartels:1995iu,Bartels:1996wc} employing the infrared evolution equations (IREE) approach \cite{Gorshkov:1966ht,Kirschner:1983di,Kirschner:1994rq,Kirschner:1994vc,Griffiths:1999dj}. Those works led to phenomenology developed in \cite{Blumlein:1995jp,Blumlein:1996hb,Ermolaev:1999jx,Ermolaev:2000sg,Ermolaev:2003zx,Ermolaev:2009cq} and allowed one to obtain predictions for the small-$x$ anomalous dimensions of the spin-dependent DGLAP evolution to higher (and potentially to all) orders in the strong coupling $\as$ \cite{Blumlein:1995jp,Blumlein:1996hb}. 

In the past decade, the question of theoretical understanding of helicity distributions at small $x$ received renewed attention \cite{Kovchegov:2015pbl, Hatta:2016aoc, Kovchegov:2016zex, Kovchegov:2016weo, Kovchegov:2017jxc, Kovchegov:2017lsr, Kovchegov:2018znm, Kovchegov:2019rrz, Boussarie:2019icw, Cougoulic:2019aja, Kovchegov:2020hgb, Cougoulic:2020tbc, Chirilli:2021lif, Adamiak:2021ppq, Kovchegov:2021lvz, Cougoulic:2022gbk}. This is due in part to the prior development of new small-$x$ resummation techniques \cite{Mueller:1994rr,Mueller:1994jq,Mueller:1995gb,Balitsky:1995ub,Balitsky:1998ya,Kovchegov:1999yj,Kovchegov:1999ua,Jalilian-Marian:1997dw,Jalilian-Marian:1997gr,Weigert:2000gi,Iancu:2001ad,Iancu:2000hn,Ferreiro:2001qy} (see \cite{Gribov:1984tu,Iancu:2003xm,Weigert:2005us,JalilianMarian:2005jf,Gelis:2010nm,Albacete:2014fwa,Kovchegov:2012mbw,Morreale:2021pnn} for reviews) which have more recently been extended to sub-eikonal (and sub-sub-eikonal) observables \cite{Altinoluk:2014oxa,Balitsky:2015qba,Balitsky:2016dgz, Kovchegov:2017lsr, Kovchegov:2018znm, Chirilli:2018kkw, Jalilian-Marian:2018iui, Jalilian-Marian:2019kaf, Altinoluk:2020oyd, Kovchegov:2021iyc, Altinoluk:2021lvu, Kovchegov:2022kyy, Altinoluk:2022jkk, Altinoluk:2023qfr} such as helicity, and in part in preparation for the data to be reported by the upcoming EIC. Novel small-$x$ evolution equations for the so-called ``polarized dipole amplitudes", which determine hPDFs and the $g_1$ structure function at small $x$, have been constructed in \cite{Kovchegov:2015pbl, Kovchegov:2016zex, Kovchegov:2017lsr, Kovchegov:2018znm} (KPS) (see also \cite{Chirilli:2021lif}). The equations resum powers of $\as \, \ln^2 (1/x)$: this is usually referred to as the double-logarithmic approximation (DLA). Important corrections modifying the KPS equations have recently been found in \cite{Cougoulic:2022gbk} (henceforth referred to as the KPS-CTT equations) using both the background field method and the light-cone operator treatment (LCOT) approach. The resulting equations have been cross-checked against the small-$x$ and large-$N_c$ part of the $\Delta \gamma_{GG}$ anomalous dimension to the three known loops \cite{Altarelli:1977zs,Dokshitzer:1977sg,Mertig:1995ny,Moch:2014sna}, indicating a complete agreement with the existing fixed-order calculations (see also \cite{Zijlstra:1993sh,Moch:1999eb,vanNeerven:2000uj,Vermaseren:2005qc,Blumlein:2021ryt,Blumlein:2021lmf,Davies:2022ofz,Blumlein:2022gpp} for other relevant calculations which, apart from presenting interesting and important results, may also be used for further cross checks of the small-$x$ resummation). Here and below $N_c$ denotes the number of quark colors.

Similarly to the unpolarized non-linear small-$x$ evolution \cite{Balitsky:1995ub,Balitsky:1998ya,Kovchegov:1999yj,Kovchegov:1999ua,Jalilian-Marian:1997dw,Jalilian-Marian:1997gr,Weigert:2000gi,Iancu:2001ad,Iancu:2000hn,Ferreiro:2001qy}, the KPS-CTT evolution gives an infinite hierarchy of equations. In the large-$N_c$ and large-$N_c \& N_f$ limits the hierarchy gets replaced by a closed system of equations \cite{Kovchegov:2015pbl, Cougoulic:2022gbk} ($N_f$ is the number of quark flavors). A numerical solution of the large-$N_c$ version of the KPS-CTT equations, performed in \cite{Cougoulic:2022gbk}, resulted in the following small-$x$ asymptotics of the flavor-singlet hPDFs and the $g_1$ structure function,
\begin{align}\label{asympt_num}
        \Delta \Sigma (x, Q^2) \sim \Delta G (x, Q^2) \sim g_1 (x, Q^2) \sim \left( \frac{1}{x} \right)^{3.66 \, \sqrt{\bas}} 
\end{align}
with
\begin{align}\label{bas_def}
    \bas \equiv \frac{\as \, N_c}{2 \pi} .
\end{align}
This result was in agreement with the small-$x$ asymptotics for hPDFs found earlier by BER in \cite{Bartels:1996wc}, with the corresponding power of $3.66 \, \sqrt{\bas}$ (the intercept) appearing to be the same as that found in \cite{Bartels:1996wc}, at least within the precision of the numerical solution performed in \cite{Cougoulic:2022gbk}.

In this paper we construct an analytic solution of the large-$N_c$ version of the KPS-CTT equations \cite{Kovchegov:2015pbl, Cougoulic:2022gbk}, that is, of the same equations which were solved numerically in \cite{Cougoulic:2022gbk} leading to the asymptotics \eqref{asympt_num}. The aims are to achieve a better understanding of these equations, obtain an analytic expression for the intercept (the power) in \eq{asympt_num}, and also perform a more detailed cross-check against the BER results \cite{Bartels:1996wc}. 

The paper is structured as follows. We state the equations we are going to solve in Sec.~\ref{sec:equations}. As mentioned above, the equations involve the ``polarized dipole amplitudes," defined in terms of operators in \cite{Cougoulic:2022gbk} (see also \cite{Kovchegov:2018znm}). We also list in Sec.~\ref{sec:equations} the relations between the polarized dipole amplitudes and hPDFs, $g_1$ structure function, and transverse momentum-dependent helicity PDFs (TMD hPDFs or hTMDs). Our analytic solution of the large-$N_c$ KPS-CTT equations is presented in Sec.~\eqref{sec:solution} and is based on a double Laplace transform method. The final results for the solution of the large-$N_c$ equations are summarized in Sec.~\ref{sec:results}, in which we also derive an analytic version of the small-$x$ asymptotics \eqref{asympt_num}, 
obtaining
 \begin{align}\label{asymptotics_general}
    \Delta \Sigma (x, Q^2) \sim \Delta G (x, Q^2) \sim g_1 (x, Q^2) \sim \left( \frac{1}{x} \right)^{\alpha_h}
\end{align}
with 
\begin{align}\label{analytic_intercept}
   \alpha_h = \frac{4}{3^{1/3}} \, \sqrt{\textrm{Re} \left[ \left( - 9 + i \, \sqrt{111} \right)^{1/3} \right] } \,\sqrt{\bas} \approx 3.66074 \, \sqrt{\bas}\,.
\end{align}
This is our exact analytic expression for the power of $3.66 \, \sqrt{\bas}$ in \eq{asympt_num}, previously obtained numerically in \cite{Cougoulic:2022gbk}. After running several cross-checks of our solution in Sec.~\ref{sec:cross_checks} and obtaining the resummed small-$x$ and large-$N_c$ anomalous dimension $\Delta \gamma_{GG} (\omega)$, we proceed by comparing our results to BER  \cite{Bartels:1996wc} in Sec.~\ref{sec:comp_BER}. There we find that the BER intercept is (cf. \cite{Kovchegov:2016zex,Cougoulic:2022gbk})
\begin{align}\label{BER_intercept}
    \alpha_h^{BER} = \sqrt{\frac{17 + \sqrt{97}}{2}} \, \sqrt{\bas} \approx 3.66394 \, \sqrt{\bas} . 
\end{align}
We see that, despite the numerical closeness of the two results \eqref{analytic_intercept} and \eqref{BER_intercept}, our intercept and that of BER are in fact different, albeit by a very small amount. The difference was not detected by the numerical solution from \cite{Cougoulic:2022gbk}: while the numerical intercept in \cite{Cougoulic:2022gbk} appeared to be closer to $3.661 \, \sqrt{\bas}$ than to the BER intercept (that is, closer to the number in \eq{analytic_intercept} than to the number in \eq{BER_intercept}), the numerical precision did not allow the authors of \cite{Cougoulic:2022gbk} to make a definitive conclusion about the difference between the numerical solution and that of BER.\footnote{Note that the earlier KPS intercept obtained in \cite{Kovchegov:2016weo, Kovchegov:2017jxc} for the un-corrected evolution differed from the BER one by about 30$\%$: compared to that, the difference between \eqref{analytic_intercept} and \eqref{BER_intercept} is rather minor.} 

A similar difference persists in the resummed small-$x$ and large-$N_c$ gluon-gluon polarized anomalous dimension $\Delta \gamma_{GG} (\omega)$, with the BER one (given by \eq{BER_anom_dim} below) being different from ours (obtained in Sec.~\ref{sec:cross_checks} and shown in \eq{anomalous_dim}). Remarkably, the expansion of the BER anomalous dimension in powers of $\as$, presented in \eq{BER_exp}, agrees with the expansion of our anomalous dimension in \eq{anomalous_dim_exp} in the first three terms, both of them agreeing with the known results for this quantity, calculated up to three loops \cite{Altarelli:1977zs,Dokshitzer:1977sg,Mertig:1995ny,Moch:2014sna}. The (rather minor) difference between the BER anomalous dimension and ours arises at order $\as^4$, that is, at the four-loop level, which has not yet been calculated in the fixed-order framework. We speculate on the origin of this minor disagreement in Appendix~\ref{A} and conclude in Sec.~\ref{sec:conc}.


\section{Large-$N_c$ Equations}
\label{sec:equations}

The large-$N_c$ DLA helicity evolution equations derived in \cite{Kovchegov:2015pbl, Kovchegov:2018znm, Cougoulic:2022gbk} are written for the (impact-parameter integrated) polarized dipole amplitudes $G (\xoz^2,zs)$ and $G_2 (\xoz^2,zs)$. These amplitudes are defined in terms of sub-eikonal operators and light-cone Wilson lines in \cite{Cougoulic:2022gbk}. The amplitudes depend on the transverse size squared of the dipole $x_{ij}^2 = |{\un x}_{ij}|^2$ for $i,j = 0,1,2, \ldots$ labeling the partons and with ${\un x}_{ij} = {\un x}_i - {\un x}_j$ for the two-dimensional transverse vectors ${\un x} = (x^1, x^2)$ in the coordinate space. The impact parameter is integrated out in these dipole amplitudes. The amplitudes also depend on the center of mass energy squared $s$ between the original projectile and the target multiplied by the smallest longitudinal momentum fraction $z$ among the two partons making up the dipole. (It is better to think of $z$ as the parameter controlling the center of mass energy squared $z s$ involved in the next step of the dipole evolution. Sometimes $z$ could be smaller than the longitudinal momentum fractions of the partons making up the dipole \cite{Kovchegov:2021lvz}, for instance, after a step of evolution involving a virtual correction.) The two dipole amplitudes $G$ and $G_2$ are accompanied by the two auxiliary (impact-parameter integrated) amplitudes $\Gamma(\xoz^2,\xto^2,z s)$ and $\Gamma_2(\xoz^2,\xto^2,z s)$, which also depend on the size squared of the adjacent dipole $\xto^2$: such amplitudes were dubbed the ``neighbor dipole amplitudes" in \cite{Kovchegov:2015pbl}. Their operator definitions are identical to those for $G$ and $G_2$, except for a difference in the light-cone lifetime cutoff \cite{Cougoulic:2019aja}, which for $\Gamma$ and $\Gamma_2$ depends on the adjacent dipole size \cite{Kovchegov:2015pbl, Kovchegov:2018znm, Cougoulic:2022gbk}. As we will see shortly below, the observables and distribution functions depend only on $G$ and $G_2$ and do not depend on the neighbor dipole amplitudes directly, such that these latter amplitudes indeed play the role of auxiliary functions present only in the evolution equations. 

The large-$N_c$ helicity evolution equations for the polarized dipole amplitudes read \cite{Kovchegov:2015pbl, Kovchegov:2018znm, Cougoulic:2022gbk}
\begin{subequations}\label{largeNc_eqns_unscaled}
\begin{align}
    \label{evolG_unscaled}
    & G(\xoz^2,zs) = G^{(0)}(\xoz^2,zs) + \frac{\as N_c}{2\pi}\int\limits_{\tfrac{1}{s\xoz^2}}^{z}\frac{\mathrm{d}z'}{z'}\int\limits_{\tfrac{1}{z's}}^{\xoz^2}\frac{\mathrm{d}\xto^2}{\xto^2}\Bigg[\Gamma(\xoz^2,\xto^2,z's) + 3 \, G(\xto^2,z's) \\
    & \hspace*{8cm} + 2 \, G_2(\xto^2,z's) + 2 \, \Gamma_2(\xoz^2,\xto^2,z's)\Bigg], \notag \\
    \label{evolGamma_unscaled}
    & \Gamma(\xoz^2,\xto^2,z's) = G^{(0)}(\xoz^2,z's) + \frac{\as N_c}{2\pi}\int\limits_{\tfrac{1}{s\xoz^2}}^{z'}\frac{\mathrm{d}z''}{z''}\int\limits_{\tfrac{1}{z''s}}^{\min\left[\xoz^2,\xto^2\tfrac{z'}{z''} \right] }\frac{\mathrm{d}\xtt^2}{\xtt^2}\Bigg[\Gamma(\xoz^2,\xtt^2,z''s) + 3 \, G(\xtt^2,z''s) \notag \\
    & \hspace*{8cm} + 2 \, G_2(\xtt^2,z''s) + 2 \, \Gamma_2(\xoz^2,\xtt^2,z''s)\Bigg],  \\
    \label{evolG2_unscaled}
    & G_2(\xoz^2,zs) = G_2^{(0)}(\xoz^2,zs) + \frac{\as N_c}{\pi}\int\limits_{\tfrac{\Lambda^2}{s}}^{z}\frac{\mathrm{d}z'}{z'}\int\limits_{\max\left[\xoz^2,\tfrac{1}{z's}\right]}^{\min\left[\tfrac{z}{z'}\xoz^2, \tfrac{1}{\Lambda^2}\right]}\frac{\mathrm{d}\xto^2}{\xto^2}\left[G(\xto^2,z's) + 2 \, G_2(\xto^2,z's) \right], \\
    \label{evolGamma2_unscaled}
    & \Gamma_2(\xoz^2,\xto^2,z's) =  G_2^{(0)}(\xoz^2,z's)  + \frac{\as N_c}{\pi}\int\limits_{\tfrac{\Lambda^2}{s}}^{z'\tfrac{\xto^2}{\xoz^2}}\frac{\mathrm{d}z''}{z''}\int\limits_{\max\left[\xoz^2,\tfrac{1}{z''s}\right]}^{\min\left[\tfrac{z'}{z''}\xto^2, \tfrac{1}{\Lambda^2}\right]}\frac{\mathrm{d}\xtt^2}{\xtt^2}\left[G(\xtt^2,z''s) + 2 \, G_2(\xtt^2,z''s) \right], 
\end{align}
\end{subequations}
where $\Gamma(\xoz^2,\xto^2,z's)$ and $\Gamma_2(\xoz^2,\xto^2,z's)$ are only defined for $\xoz\geq\xto$ and $\Lambda$ is an infrared (IR) cutoff such that we require all the dipole sizes to be $x_{ij} <1/\Lambda$.

For convenience, we define the new variables \cite{Kovchegov:2016weo}
\begin{gather}
\label{originalvars}
    \eta = \sqrt{\frac{\as N_c}{2\pi}}\ln\frac{zs}{\Lambda^2}\,, \quad \eta' = \sqrt{\frac{\as N_c}{2\pi}}\ln\frac{z's}{\Lambda^2}\,, \quad \eta'' = \sqrt{\frac{\as N_c}{2\pi}}\ln\frac{z''s}{\Lambda^2}\,, \\ 
    \soz = \sqrt{\frac{\as N_c}{2\pi}} \ln\frac{1}{x_{10}^2\Lambda^2}\,, \quad \sto = \sqrt{\frac{\as N_c}{2\pi}} \ln\frac{1}{x_{21}^2\Lambda^2}\,, \quad \stt = \sqrt{\frac{\as N_c}{2\pi}} \ln\frac{1}{x_{32}^2\Lambda^2}\,.\notag
\end{gather}
In terms of these, \eqs{largeNc_eqns_unscaled} can be written as
\begin{subequations}\label{largeNc_eqns}
\begin{align}
\label{evolG}
& G(\soz,\eta) = G^{(0)}(\soz,\eta) + \int\limits_{\soz}^{\eta}\mathrm{d}\eta' \int\limits_{\soz}^{\eta'}\mathrm{d}\sto \, \bigg[ \Gamma(\soz,\sto,\eta') + 3G(\sto,\eta') + 2 \, G_2(s_{21},\eta') + 2 \, \Gamma_2(s_{10},s_{21},\eta')\bigg],  \\
\label{evolGamma}
& \Gamma(\soz,\sto,\eta') = G^{(0)}(\soz,\eta') + \Bigg[ \ \int\limits_{\soz}^{\sto}\mathrm{d} \stt \int\limits_{\stt}^{\eta'-\sto+\stt}\mathrm{d}\eta'' + \int\limits_{\sto}^{\eta'}\mathrm{d}\stt \int\limits_{\stt}^{\eta'}\mathrm{d}\eta''  \Bigg] \\ 
& \hspace*{5cm} \times\bigg[ \Gamma(\soz,\stt,\eta'') + 3 \, G(\stt,\eta'') + 2 \, G_2(\stt,\eta'') + 2 \, \Gamma_2(\soz,\stt,\eta'')\bigg], \notag \\
\label{evolG2}
& G_2(\soz,\eta) = G_2^{(0)}(\soz,\eta) + 2 \int\limits_0^{\soz}\mathrm{d}\sto \int\limits_{\sto}^{\eta-\soz+\sto}\mathrm{d}\eta' \bigg[G(\sto,\eta') + 2 \, G_2(\sto,\eta')\bigg], \\
\label{evolGamma2}
& \Gamma_2(\soz,\sto,\eta') = G_2^{(0)}(\soz,\eta') + 2 \int\limits_0^{\soz}\mathrm{d}\stt \int\limits_{\stt}^{\eta'-\sto+\stt}\mathrm{d}\eta'' \bigg[G(\stt,\eta'') + 2 \, G_2(\stt,\eta'')\bigg],
\end{align}
\end{subequations}
where we have changed the order of integration in the integral kernels of Eqs. \eqref{evolGamma_unscaled}-\eqref{evolGamma2_unscaled}. Once again, the ordering $0\leq\soz\leq\sto\leq\eta'$ is assumed in Eqs. \eqref{evolGamma} and \eqref{evolGamma2}.

Once the dipole amplitudes $G$ and $G_2$ are determined by solving Eqs.~\eqref{largeNc_eqns}, they can be used to calculate 
the (dipole) gluon and (flavor-singlet) quark helicity TMDs $g^{G \, dip}_{1L}(x,k_T^2)$ and $g^{S}_{1L}(x,k_T^2)$, 
hPDFs $\Delta G(x,Q^2)$ and $\Delta \Sigma(x,Q^2)$, and the $g_1$ structure function, by employing the following relations derived in \cite{Cougoulic:2022gbk} (see also \cite{Kovchegov:2015pbl, Kovchegov:2017lsr, Kovchegov:2018znm}):
\begin{subequations}\label{distributions+g1}
\begin{align}
    \label{gluon_TMD}
    & g^{G\,dip}_{1L}(x,k_T^2) = \frac{N_c}{\as 2\pi^4}\int \mathrm{d}^2 \xoz\, e^{-i\underline{k}\cdot\underline{x}_{10}} \left[1+\xoz^2\frac{\partial}{\partial\xoz^2}\right]G_2\left(\xoz^2,zs=\frac{Q^2}{x}\right),\\
    \label{quark_TMD}
    & g^{S}_{1L}(x,k_T^2) = \frac{8iN_cN_f}{(2\pi)^5}\int\limits_{\Lambda^2/s}^{1}\frac{\mathrm{d}z}{z}\int \mathrm{d}^2 \xoz\, e^{i\underline{k}\cdot\underline{x}_{10}} \, \frac{\underline{x}_{10}}{\xoz^2}\cdot \frac{\underline{k}}{\underline{k}^2}\left[Q(\xoz^2,zs) + 2 \, G_2(\xoz^2,zs)\right],\\
    \label{gluon_PDF}
    & \Delta G(x,Q^2) = \frac{2 N_c}{\as \pi^2}\left[\left(1+\xoz^2\frac{\partial}{\partial \xoz^2} \right) G_2\left(\xoz^2, zs=\frac{Q^2}{x} \right) \right]_{\xoz^2 = \tfrac{1}{Q^2}}, \\
    \label{quark_PDF}
    & \Delta \Sigma(x,Q^2) = -\frac{N_cN_f}{2\pi^3}\int\limits_{\Lambda^2/s}^{1}\frac{\mathrm{d}z}{z}\int\limits_{\tfrac{1}{zs}}^{\min\left\{\tfrac{1}{zQ^2},\tfrac{1}{\Lambda^2} \right\}} \frac{\mathrm{d}\xoz^2}{\xoz^2}\left[Q(\xoz^2,zs) + 2 \, G_2(\xoz^2,zs)\right],\\
    \label{g1}
    & g_1(x,Q^2) = -\sum_{f}\frac{N_c Z_f^2}{4\pi^3}\int\limits_{\Lambda^2/s}^{1}\frac{\mathrm{d}z}{z} \int\limits_{\tfrac{1}{zs}}^{\min\left\{\tfrac{1}{zQ^2}, \tfrac{1}{\Lambda^2} \right\}} \frac{\mathrm{d}\xoz^2}{\xoz^2}\left[Q(\xoz^2,zs) + 2 \, G_2(\xoz^2,zs) \right].
\end{align}
\end{subequations}
The transverse momentum vector is denoted by ${\un k} = (k^1, k^2)$ and its magnitude is $k_T = |{\un k}|$. We have introduced the fractional electric charge of the quark $Z_f$. In \eq{quark_PDF} we have also assumed, for simplicity, that all flavors contribute equally, such that the sum over flavors can be replaced by the number of flavors $N_f$. This appears to be a good approximation at large $N_c$. In addition, we have employed another polarized dipole amplitude $Q(\xoz^2,zs)$, which is also defined in \cite{Cougoulic:2022gbk}. Note that at large $N_c$, one has $Q(\xoz^2,zs) \approx G(\xoz^2,zs)$ \cite{Cougoulic:2022gbk}, such that all the quantities in Eqs.~\eqref{distributions+g1} can be expressed in terms of the amplitudes $G$ and $G_2$.


\section{Solution}

\label{sec:solution}


\subsection{Double Inverse Laplace Transform Representations for $G_2,\,\Gamma_2,\,G$}

Our goal now is to solve Eqs.~\eqref{largeNc_eqns}. We begin by writing $G_2(\soz,\eta)$ as a double inverse Laplace transform over the variables $\eta-\soz$ and $\soz$:
\begin{align}
\label{G2}
    G_2(\soz,\eta) = \wint \gint \, e^{\omega(\eta-\soz)}e^{\gamma\soz}\gtwg.
\end{align}
The integrals here are taken over infinite straight-line contours in the complex $\omega$- and $\gamma$-planes, parallel to the imaginary axis and to the right of all the integrand's singularities.

We can also introduce corresponding double inverse Laplace transforms for the initial conditions/inhomogeneous terms $G^{(0)}(\soz,\eta)$ and $G_2^{(0)}(\soz,\eta)$,
\begin{subequations}\label{initial_conditions}
\begin{align}
    \label{G0}
    & G^{(0)}(\soz,\eta) = \wint \gint \, e^{\omega(\eta-\soz)}e^{\gamma\soz}G_{\omega\gamma}^{(0)}\,,\\
    \label{G20}
    & G_2^{(0)}(\soz,\eta) = \wint \gint \, e^{\omega(\eta-\soz)}e^{\gamma\soz}G_{2\omega\gamma}^{(0)}\,.
\end{align}
\end{subequations}

Next we observe that Eqs. \eqref{evolG2} and \eqref{evolGamma2} admit the following scaling property:
\begin{align}
\label{G2Gamma2scaling}
    \Gamma_2(\soz,\sto,\eta') - G_2^{(0)}(\soz,\eta') = G_2(\soz,\eta=\eta'+\soz-\sto) - G_2^{(0)}(\soz,\eta=\eta'+\soz-\sto).
\end{align}
Using Eqs. \eqref{G2} and \eqref{G20} in \eq{G2Gamma2scaling} we immediately have
\begin{align}
\label{Gamma2}
    \Gamma_2(\soz,\sto,\eta') = \wint\gint \left[e^{\omega(\eta'-\sto)}e^{\gamma\soz}\left(\gtwg - \gtwg^{(0)}\right) + e^{\omega(\eta'-\soz)}e^{\gamma\soz} \, \gtwg^{(0)} \right].
\end{align}

Now we write the amplitude $G(\soz,\eta)$ as a double inverse Laplace transform
\begin{align}
\label{Gprelim}
    G(\soz,\eta) = \wint \gint \, e^{\omega(\eta-\soz)}e^{\gamma\soz}G_{\omega\gamma}\,,
\end{align}
substitute Eqs.~\eqref{G2}, \eqref{G20}, and \eqref{Gprelim} into \eqref{evolG2} and perform the integrals over $\eta'$ and $\sto$. Next, applying the forward Laplace transforms over $\eta-\soz$ and $\soz$ (treating those as two independent variables) yields

\begin{align}\label{c3}
    G_{2 \omega \gamma} = G_{2 \omega \gamma}^{(0)} + \frac{2}{\omega \gamma} \left[G_{\omega\gamma} + 2 \, G_{2 \omega \gamma} \right].
\end{align}
Solving \eq{c3} for $G_{\omega\gamma}$ we arrive at
\begin{align}
\label{G_omegagamma_0}
    G_{\omega\gamma} = \tfrac{1}{2}\omega\gamma\left(\gtwg - \gtwg^{(0)} \right) - 2\gtwg\,,
\end{align}
so that \eq{Gprelim} gives
\begin{align}
\label{G}
    G(\soz,\eta) = \wint \gint e^{\omega(\eta-\soz)}e^{\gamma\soz}\left[\tfrac{1}{2}\omega\gamma\left(\gtwg-\gtwg^{(0)} \right) - 2\gtwg \right].
\end{align}
This way, we have obtained double inverse Laplace transform representations for the dipole amplitudes $G_2, \Gamma_2$ and $G$ given in Eqs.~\eqref{G2}, \eqref{Gamma2} and \eqref{G}, respectively.

From Eqs. \eqref{evolG2} and \eqref{evolGamma2} we have several boundary conditions which our expressions for $G_2(\soz,\eta)$ and $\Gamma_2(\soz,\sto,\eta')$ must satisfy. We need
\begin{subequations}
\begin{align}
    \label{G2_bc0)}
    G_2(\soz=0,\eta) &= G_2^{(0)}(\soz=0,\eta)\,, \\
    \label{G2_bc1}
    G_2(\soz,\eta=\soz) &= G_2^{(0)}(\soz,\eta=\soz)\,, \\
    \label{Gamma2_bc0}
    \Gamma_2(\soz=0,\sto,\eta') &= G_2^{(0)}(\soz=0,\eta')\,, \\
    \label{Gamma2_bc1}
    \Gamma_2(\soz,\sto,\eta'=\sto) &= G_2^{(0)}(\soz,\eta'=\sto)\,.
\end{align}
\end{subequations}
Using Eqs. \eqref{G2} and \eqref{G20}, we see that Eqs. \eqref{G2_bc0)} and \eqref{G2_bc1} give respectively
\begin{subequations}\label{G2_bcs_laplace}
\begin{align}
    \label{G2_bcs_laplace_0}
    \wint \gint e^{\omega\eta}\gtwg &= \wint \gint e^{\omega\eta}\gtwg^{(0)}\,,\\
    \label{G2_bcs_laplace_1}
    \wint \gint e^{\gamma\soz}\gtwg &= \wint \gint e^{\gamma\soz}\gtwg^{(0)}. 
\end{align}
\end{subequations}
Note that the constraints resulting from Eqs. \eqref{Gamma2_bc0} and \eqref{Gamma2_bc1} are equivalent to Eqs. \eqref{G2_bcs_laplace}. 

Since $\gtwg, G_{\omega\gamma}$ must go to zero as $\omega\rightarrow\infty$ or $\gamma\rightarrow\infty$ in order for the Laplace transforms to exist, the second term on the right-hand side of \eq{c3} goes to zero faster than $1/\omega$ or $1/\gamma$ as $\omega\rightarrow\infty$ or $\gamma\rightarrow\infty$. This implies that
\begin{align}
    \int \frac{d \omega}{2\pi i} \, \frac{2}{\omega \gamma} \left[G_{\omega\gamma} + 2 \, G_{2 \omega \gamma} \right] = 0
\end{align}
and
\begin{align}
    \int \frac{d \gamma}{2\pi i} \, \frac{2}{\omega \gamma} \left[G_{\omega\gamma} + 2 \, G_{2 \omega \gamma} \right] = 0,
\end{align}
since the $\omega$- and $\gamma$-contours are located to the right of all the singularities of the integrand, allowing us to close the contours to the right. We see that the conditions in Eqs.~\eqref{G2_bcs_laplace} are automatically satisfied by \eq{c3}.

We conclude that at this point Eqs.~\eqref{evolG2} and \eqref{evolGamma2} are completely solved. 


\subsection{Double Inverse Laplace Transform Representation for $\Gamma$}

Differentiating \eq{evolGamma} one can show that $\Gamma(\soz,\sto,\eta')$ satisfies the partial differential equation
\begin{align}
\label{Gamma_pde}
    \frac{\partial^2\Gamma(\soz,\sto,\eta')}{\partial \sto^2} +  \frac{\partial^2\Gamma(\soz,\sto,\eta')}{\partial \sto \partial \eta'} + \Gamma(\soz,\sto,\eta') = -3 \, G(\sto,\eta') - 2 \, G_2(\sto,\eta') - 2\, \Gamma_2(\soz,\sto,\eta').
\end{align}
This second-order partial differential equation has two solutions, homogeneous and particular, which we label (h) and (p), respectively,
\begin{align}\label{Gamma_sol1}
    \Gamma(\soz,\sto,\eta') = \Gamma^{(h)}(\soz,\sto,\eta') + \Gamma^{(p)}(\soz,\sto,\eta') .
\end{align}

Looking for the homogeneous solution of the form 
\begin{align}
\Gamma^{(h)}(\soz,\sto,\eta')  = \wint\gint e^{\omega(\eta'-\sto)}e^{\gamma\sto}\Gamma_{\omega\gamma} (\soz)
\end{align}
one arrives at the condition 
\begin{align}
\gamma^2 - \omega\gamma + 1 = 0 ,
\end{align}
which yields two solutions, $\gamma = \delta^+_\omega$ and $\gamma = \delta^-_\omega$, where we have defined
\begin{align}\label{delta_omega_pm}
   \dw^\pm \equiv \frac{\omega}{2}\left[1\pm\sqrt{1-\tfrac{4}{\omega^2}} \right].
\end{align}
Thus, the homogeneous solution can be written as
\begin{align}
\label{HomogeneousGammaSimple2}
    \Gamma^{(h)}(\soz,\sto,\eta') = \wint \, e^{\omega(\eta'-\sto)}\big[ \Gamma_{\omega}^+(\soz) \, e^{\dw^+\sto} + \Gamma_{\omega}^-(\soz) \, e^{\dw^-\sto}   \big] 
\end{align}
with some unknown functions $\Gamma_{\omega}^+(\soz)$ and $\Gamma_{\omega}^-(\soz)$.

To construct a particular solution of \eq{Gamma_pde}, one can substitute Eqs. \eqref{G}, \eqref{G2}, and \eqref{Gamma2} into the right hand side of \eq{Gamma_pde}. This motivates an ansatz for the particular solution of the form 
\begin{align}
\label{Gamma_particular_ansatz}
    \Gamma^{(p)}(\soz,\sto,\eta') = \wint \gint \bigg[A_{\omega\gamma} \, e^{\omega(\eta'-\sto)}e^{\gamma\sto} + B_{\omega\gamma} \, e^{\omega(\eta'-\sto)}e^{\gamma\soz} + C_{\omega\gamma} \, e^{\omega(\eta'-\soz)}e^{\gamma\soz}\bigg].
\end{align}
Substitution of \eq{Gamma_particular_ansatz} into \eq{Gamma_pde} allows one to determine the coefficients $A_{\omega\gamma}$, $B_{\omega\gamma}$, and $C_{\omega\gamma}$. The particular solution found this way is 
\begin{align}
\label{GammaParticularSimple}
        \Gamma^{(p)}(\soz,\sto,\eta') = \wint \gint \bigg[&e^{\omega(\eta'-\sto)}e^{\gamma\sto}\bigg(\stf \gtwg + 
        \frac{\frac{3}{2}\omega\gamma}{\gamma^2-\omega\gamma+1}\gtwg^{(0)} \bigg) \\
        &-2 \, e^{\omega(\eta'-\sto)}e^{\gamma\soz}\big[\gtwg - \gtwg^{(0)} \big]
        -2 \, e^{\omega(\eta'-\soz)}e^{\gamma\soz}\gtwg^{(0)} \bigg] . \notag
\end{align}

Combining the homogeneous \eqref{HomogeneousGammaSimple2} and particular \eqref{GammaParticularSimple} solutions we arrive at the general solution of \eq{Gamma_pde}, 
\begin{align}\label{Gamma}
    \Gamma(\soz,\sto,\eta') = &\wint e^{\omega(\eta'-\sto)}\left[\Gamma_{\omega}^+(\soz)e^{\dw^+\sto} + \Gamma_{\omega}^-(\soz)e^{\dw^-\sto} \right] \\
    +&\wint\gint \Bigg[e^{\omega(\eta'-\sto)}e^{\gamma\sto}\left(\frac{- \tfrac{3}{2}\omega\gamma + 4}{\gamma^2 - \omega\gamma + 1}\gtwg + \frac{\tfrac{3}{2}\omega\gamma}{\gamma^2 - \omega\gamma + 1}\gtwg^{(0)}  \right)\notag \\
    &\qquad\qquad\qquad -2 \, e^{\omega(\eta'-\sto)}e^{\gamma\soz}\left(\gtwg - \gtwg^{(0)}\right) - 2 \, e^{\omega(\eta'-\soz)}e^{\gamma\soz}\gtwg^{(0)}\Bigg]\notag.
\end{align}

The integral in \eq{Gamma} is not defined until we specify the location of the poles in the new denominator, $\gamma^2 - \omega\gamma + 1 = -\gamma[\omega-(\gamma+\tfrac{1}{\gamma}) ] = (\gamma-\dw^+)(\gamma-\dw^-)$, with respect to the $\omega$- and $\gamma$-contours. A simple analysis shows that one cannot have both the $\omega$- and $\gamma$-contours to the right of the new poles. Indeed, since
\begin{align}
    \gamma^2-\omega\gamma+1 = (\gamma - \delta_\omega^+) \, (\gamma - \delta_\omega^-) \stackrel{|\omega| \to \infty}{\longrightarrow} (\gamma - \omega) \, \left( \gamma - \frac{1}{\omega} \right),
\end{align}
we see that if Re~$\omega > $~Re~$\gamma$ then the $\gamma = \omega$ pole is to the right of the $\gamma$-contour and to the left of the $\omega$-contour; if Re~$\omega <$~Re~$\gamma$ than the $\gamma = \omega$ pole is to the left of the $\gamma$-contour and to the right of the $\omega$-contour.  
We choose the $\omega$-contour to be to the right of the singularity at $\omega = \gamma + 1/\gamma$ generated by the new denominator. Then, as one can show, the $\gamma$-contour must pass between the $\gamma = \dw^+$ and $\gamma = \dw^-$ poles. We stress that the locations of $\omega$- and $\gamma$-contours here are a choice, affecting both the homogeneous and particular solutions simultaneously: different choices for the contours' locations would result in different $\Gamma_{\omega}^+(\soz)$ and $\Gamma_{\omega}^-(\soz)$. As we will see below, the residue at $\gamma=\dw^+$ will be zero in the final solution. Therefore, this new pole to the right of the $\gamma$-contour will vanish, such that all the $\gamma$-singularities of the integrand will still be to the left of the $\gamma$-contour, as expected for an inverse Laplace transform. However, we will need to keep this pole in mind later when we invert an integral over $\gamma$.


\subsection{Constraints on $\Gamma$}

Note that we might have lost some of the constraints of \eq{evolGamma} when we differentiated it to obtain \eq{Gamma_pde}. Hence, the expression \eqref{Gamma}, while a solution of the differential equation \eqref{Gamma_pde}, may not yet be a solution of \eq{evolGamma}. To fully satisfy \eq{evolGamma} we take our expression \eqref{Gamma} for $\Gamma(\soz,\sto,\eta')$ along with the other three amplitudes given in Eqs. \eqref{G2}, \eqref{Gamma2}, and \eqref{G} and substitute them all back into \eq{evolGamma}. Performing the integrals over $\eta''$ and $\stt$ and also making use of the facts that $\dw^+\dw^- = 1$ and $\dw^+ + \dw^- = \omega$ (as can be seen from \eq{delta_omega_pm}), we obtain
\begin{align}
\label{back_sub_for_Gamma_0}
    0 = &\wint \gint e^{\omega(\eta'-\sto)}e^{\gamma\soz}\Bigg\{ \frac{\gamma-\omega}{\omega}\left[\frac{- \tfrac{3}{2}\omega\gamma + 4}{\gamma^2-\omega\gamma+1}\gtwg + \frac{\tfrac{3}{2}\omega\gamma}{\gamma^2-\omega\gamma+1}\gtwg^{(0)} \right] + 2\left(\gtwg-\gtwg^{(0)} \right) \Bigg\} \\
    &+ \wint\gint e^{\omega(\eta'-\soz)}e^{\gamma\soz}\left(G_{\omega\gamma}^{(0)} + 2\gtwg^{(0)}\right)\notag\\
    &-\wint \left\{ \Gamma_{\omega}^+(\soz)\left[ \frac{e^{\omega(\eta'-\sto)}e^{\dw^+\soz}}{\omega \, \dw^+} + e^{\dw^+\eta'} - \frac{e^{\dw^+\soz}}{\omega \, \dw^+}\right] + \Gamma_{\omega}^- (\soz)\left[ \frac{e^{\omega(\eta'-\sto)}e^{\dw^- \soz}}{\omega \, \dw^-} + e^{\dw^-\eta'} - \frac{e^{\dw^-\soz}}{\omega \, \dw^-}\right]  \right\} \notag.
\end{align}
In arriving at \eq{back_sub_for_Gamma_0} we have dropped the following term:
\begin{align}
    \wint \gint \left(\frac{e^{\gamma\eta'}}{\gamma(\gamma-\omega)} + \frac{e^{\gamma\soz}}{\omega\gamma}\right)\gamma(\gamma-\omega)\left(\frac{\tfrac{3}{2}\omega\gamma - 4}{\gamma^2-\omega\gamma+1}\gtwg - \frac{\tfrac{3}{2}\omega\gamma}{\gamma^2-\omega\gamma+1}\gtwg^{(0)} \right).
\end{align}
With no $\omega$ in the exponent in this term, we can close the $\omega$-contour to the right. Then using the fact that $\gtwg$, $\gtwg^{(0)}$ must go to zero for $\omega\rightarrow\infty$ along with the property in \eq{G2_bcs_laplace_1}, one can show that this entire term is zero.

Performing the forward Laplace transform over $\eta'$ in \eq{back_sub_for_Gamma_0} we obtain
\begin{align}
\label{back_sub_for_Gamma_1}
    0 &= e^{-\omega\sto} \, \gint \, e^{\gamma\soz}\Bigg\{ \frac{\gamma-\omega}{\omega}\left[\frac{\tfrac{-3}{2}\omega\gamma + 4}{\gamma^2-\omega\gamma+1}\gtwg + \frac{\tfrac{3}{2}\omega\gamma}{\gamma^2-\omega\gamma+1}\gtwg^{(0)} \right] + 2\left(\gtwg-\gtwg^{(0)} \right)\Bigg\} \\
    &- \Gamma_\omega^+(\soz) \frac{e^{-\omega\sto}e^{\dw^+\soz}}{\omega \, \dw^+} - \Gamma_\omega^-(\soz) \frac{e^{-\omega\sto}e^{\dw^-\soz}}{\omega \, \dw^-}\notag \\
    &+\gint e^{-\omega\soz}e^{\gamma\soz}\bigg(G_{\omega\gamma}^{(0)} + 2\gtwg^{(0)}\bigg) + \int \frac{\mathrm{d}\omega'}{2\pi i}\Bigg(\frac{\Gamma_{\omega'}^+(\soz)}{\delta_{\omega'}^+ - \omega} + \frac{\Gamma_{\omega'}^-(\soz)}{\delta_{\omega'}^- - \omega} \Bigg) \notag \\
    & + \frac{1}{\omega}\int \frac{\mathrm{d}\omega'}{2\pi i}\left(\Gamma_{\omega'}^+(\soz)\frac{e^{\delta_{\omega'}^+\soz}}{\omega' \, \delta_{\omega'}^+} + \Gamma_{\omega'}^-(\soz)\frac{e^{\delta_{\omega'}^-\soz}}{\omega' \, \delta_{\omega'}^-}\right)\notag .
\end{align}
Note that the terms in the first two lines of \eq{back_sub_for_Gamma_1} have the same $\sto$-dependence, $\propto e^{-\omega\sto}$, whereas the last two lines are independent of $\sto$. Since \eq{back_sub_for_Gamma_1} must be valid for all $\sto >0$, we conclude that the sum of the first two lines in \eq{back_sub_for_Gamma_1} must be separately equal to zero. This means that the sum of the last two lines in \eq{back_sub_for_Gamma_1} must also be zero. This gives two constraints
\begin{subequations}\label{b-cond}
\begin{align}
    \label{back_sub_for_Gamma_2}
    & \gint e^{\gamma\soz}\Bigg\{ \frac{\gamma-\omega}{\omega}\left[\frac{\tfrac{-3}{2}\omega\gamma + 4}{\gamma^2-\omega\gamma+1}\gtwg + \frac{\tfrac{3}{2}\omega\gamma}{\gamma^2-\omega\gamma+1}\gtwg^{(0)} \right] + 2\left(\gtwg-\gtwg^{(0)} \right)\Bigg\} \\
    & = \Gamma_\omega^+(\soz) \frac{e^{\dw^+\soz}}{\omega \, \dw^+} + \Gamma_\omega^-(\soz) \frac{e^{\dw^-\soz}}{\omega \, \dw^-}\,,\notag\\
    \label{back_sub_for_Gamma_3}
    & 0 = \gint e^{-\omega\soz}e^{\gamma\soz}\bigg(G_{\omega\gamma}^{(0)} + 2\gtwg^{(0)}\bigg) + \int \frac{\mathrm{d}\omega'}{2\pi i}\Bigg(\frac{\Gamma_{\omega'}^+(\soz)}{\delta_{\omega'}^+ - \omega} + \frac{\Gamma_{\omega'}^-(\soz)}{\delta_{\omega'}^- - \omega} \Bigg) \\
    & \hspace*{1cm} + \frac{1}{\omega}\int \frac{\mathrm{d}\omega'}{2\pi i}\left(\Gamma_{\omega'}^+(\soz)\frac{e^{\delta_{\omega'}^+\soz}}{\omega' \, \delta_{\omega'}^+} + \Gamma_{\omega'}^-(\soz)\frac{e^{\delta_{\omega'}^-\soz}}{\omega' \, \delta_{\omega'}^-}\right)\,.\notag
\end{align}    
\end{subequations}
If we satisfy the conditions \eqref{b-cond}, we will solve \eq{evolGamma}.

Before we do that, we observe that the evolution equation \eqref{evolG} for $G(\soz,\eta)$ is just a special case of \eq{evolGamma} for $\Gamma(\soz,\sto,\eta)$. So to ensure that \eq{evolG} is satisfied as well, we impose the condition
\begin{align}\label{G=Gamma}
\Gamma(\soz,\sto=\soz,\eta) = G(\soz,\eta), 
\end{align}
which follows from Eqs.~\eqref{evolG} and \eqref{evolGamma}. We then substitute the dipole amplitudes from Eqs.~\eqref{G} and \eqref{Gamma} into \eq{G=Gamma} and perform the forward Laplace transform over $\eta-\soz$, obtaining
\begin{align}
\label{G_Gamma_scaling}
    \gint e^{\gamma\soz}\left[ \frac{\tfrac{3}{2}\omega\gamma-4}{\gamma^2-\omega\gamma+1}\gtwg -  \frac{\tfrac{3}{2}\omega\gamma}{\gamma^2-\omega\gamma+1}\gtwg^{(0)} + \tfrac{1}{2}\omega\gamma\left(\gtwg - \gtwg^{(0)} \right)  \right] 
    = \Gamma_{\omega}^+(\soz) \, e^{\dw^+\soz} + \Gamma_{\omega}^-(\soz) \, e^{\dw^-\soz}\,.
\end{align}

Equations \eqref{back_sub_for_Gamma_2} and \eqref{G_Gamma_scaling} can be solved to give individual expressions for $\Gamma_{\omega}^{+}(\soz)$ and $\Gamma_{\omega}^{-}(\soz)$. Making use of the properties $(\dw^\pm)^2 - \omega\dw^\pm + 1 = 0$, $\dw^+\dw^- = 1$, $\dw^+ + \dw^- = \omega$, which follow from \eq{delta_omega_pm}, the result can be written as
\begin{subequations}\label{Gamma_omega_pm}
\begin{align}
    \label{Gamma_omega_plus}
    \Gamma_{\omega}^+(\soz) &= \frac{e^{-\dw^+\soz}}{\dw^+-\dw^-}\gint e^{\gamma\soz}\frac{\omega \, \dw^+}{2 \, (\gamma-\dw^+)}\left[\gtwg\left(\gamma^2-\omega\gamma+4-\tfrac{8}{\omega} \, \dw^- \right) - \gtwg^{(0)}\left(\gamma^2-\omega\gamma+4\right)  \right], \\
    \label{Gamma_omega_minus}
    \Gamma_{\omega}^-(\soz) &= \frac{e^{-\dw^-\soz}}{\dw^--\dw^+}\gint e^{\gamma\soz}\frac{\omega \, \dw^-}{2 \, (\gamma-\dw^-)}\left[\gtwg\left(\gamma^2-\omega\gamma+4-\tfrac{8}{\omega} \, \dw^+ \right) - \gtwg^{(0)}\left(\gamma^2-\omega\gamma+4\right)  \right].
\end{align}
\end{subequations}

We are only left with \eq{back_sub_for_Gamma_3} to satisfy. Employing \eq{c3} in Eqs.~\eqref{Gamma_omega_pm} along with the fact that $G_{2\omega \gamma}, G_{2\omega \gamma}^{(0)} \rightarrow 0$ as $\omega \rightarrow \infty$ we conclude that 
\begin{align}
 e^{\dw^+\soz} \, \Gamma_{\omega}^+(\soz) \to 0, \ \ \  e^{\dw^- \soz} \, \omega \, \Gamma_{\omega}^- (\soz) \to 0, \ \ \ \textrm{when} \ \ \  \omega \to \infty.
\end{align}
(We have also employed the fact that $\dw^+ \to \omega$ and $\dw^- \to 1/\omega$ as $\omega \to \infty$, which follows from \eq{delta_omega_pm}). This allows us to close the $\omega'$ contour to the right in the last term of \eq{back_sub_for_Gamma_3}, obtaining zero.  
\eq{back_sub_for_Gamma_3} then becomes
\begin{align}\label{back_sub_for_Gamma_4}
    \gint e^{-\omega\soz}e^{\gamma\soz}\bigg(G_{\omega\gamma}^{(0)} + 2 \, \gtwg^{(0)}\bigg) &= \int \frac{\mathrm{d}\omega'}{2\pi i}\Bigg(\frac{\Gamma_{\omega'}^+(\soz)}{\omega - \delta_{\omega'}^+} + \frac{\Gamma_{\omega'}^-(\soz)}{\omega - \delta_{\omega'}^-} \Bigg) \\
    &= -\int \frac{\mathrm{d}\omega'}{2\pi i}\Bigg(\frac{\omega-\delta_{\omega'}^-}{\omega}\frac{\Gamma_{\omega'}^+(\soz)}{\omega'-\left(\omega+\tfrac{1}{\omega} \right)} + \frac{\omega-\delta_{\omega'}^+}{\omega}\frac{\Gamma_{\omega'}^-(\soz)}{\omega'-\left(\omega+\tfrac{1}{\omega}\right)} \Bigg)\notag\,,
\end{align}
where we have again used the properties $\dw^+\dw^- = 1$, $\dw^+ + \dw^- = \omega$ to obtain the second line of \eq{back_sub_for_Gamma_4}. Now we can close the $\omega'$-contour to the right, picking up the pole at $\omega' = \omega + \tfrac{1}{\omega}$ in each term. Using $\delta^-_{\omega+\frac{1}{\omega}} = \frac{1}{\omega}$ and $\delta^+_{\omega+\frac{1}{\omega}} = \omega$ (as can be seen from \eq{delta_omega_pm}) we see that the $\Gamma^-_{\omega'}(\soz)$ term vanishes. Then \eq{back_sub_for_Gamma_4} has become
\begin{align}\label{back_sub_for_Gamma_5}
    \gint e^{\gamma\soz}\bigg(G_{\omega\gamma}^{(0)} + 2\gtwg^{(0)}\bigg) = \left(1-\frac{1}{\omega^2}\right) \,\Gamma^+_{\omega+\frac{1}{\omega}}\,,
\end{align}
where we have defined $\Gamma_\omega^+$ 
by $\Gamma_\omega^+(\soz) \equiv \Gamma_\omega^+ \, e^{-\dw^+\soz}$. Again using $\delta^+_{\omega + \frac{1}{\omega}} = \omega$, $\delta^-_{\omega + \frac{1}{\omega}} = 1/\omega$, we rewrite \eq{back_sub_for_Gamma_5} as
\begin{align}\label{back_sub_for_Gamma_5.3}
    \gint e^{\gamma\soz}\bigg(G_{\delta^+_{\omega + \frac{1}{\omega}} \gamma}^{(0)} + 2 \, G^{(0)}_{2 \, \delta^+_{\omega + \frac{1}{\omega}} \gamma} \bigg) = \left(1-\frac{1}{\left[ \delta^+_{\omega + \frac{1}{\omega}} \right]^2}\right) \, \Gamma^+_{\omega+\frac{1}{\omega}}\,,
\end{align}
or, replacing $\omega + \tfrac{1}{\omega} \to \omega$, as
\begin{align}\label{back_sub_for_Gamma_5.7}
    \gint e^{\gamma\soz}\bigg(G_{\delta^+_{\omega} \gamma}^{(0)} + 2 \, G^{(0)}_{2 \, \delta^+_{\omega} \gamma} \bigg) = \left(1-\frac{1}{\left[ \delta^+_{\omega} \right]^2}\right) \, \Gamma^+_{\omega}\, .
\end{align}

Employing \eq{Gamma_omega_plus} in \eq{back_sub_for_Gamma_5.7} and inverting the $\gamma$ integral --- while remembering that the $\gamma = \delta^+_\omega$ pole in the former equation is located to the right of the $\gamma$-contour --- we arrive at

\begin{align}\label{G2_omega_gamma_constraint0}
    G^{(0)}_{\dw^+\gamma} + 2 \, G^{(0)}_{2 \, \dw^+\gamma} = \frac{\omega}{2 \, (\gamma-\dw^+)}\bigg[&\gtwg\left(\gamma-\gamma^-_\omega\right)\left(\gamma-\gamma^+_\omega\right) - \gtwg^{(0)}\left(\gamma^2-\omega\gamma+4\right) \\
    &- G_{2\omega\dw^+}\left(\dw^+ - \gamma^-_\omega \right)\left(\dw^+ - \gamma^+_\omega \right) + 3 \, G^{(0)}_{2\omega\dw^+}\bigg]\notag.
\end{align}
In arriving at \eq{G2_omega_gamma_constraint0} we have again used $(\dw^+)^2 - \omega\dw^+ + 1 =0$ and have also defined
\begin{align}
\gamma^2 -\omega\gamma + 4 - \frac{8}{\omega} \, \dw^- \equiv \left(\gamma-\gamma^-_\omega\right)\left(\gamma-\gamma^+_\omega\right)
\end{align}
with
\begin{align}\label{gamma_omega_pm}
    \gamma^{\pm}_\omega = \frac{\omega}{2}\left[1 \pm \sqrt{1 - \frac{16}{\omega^2}\sqrt{1-\frac{4}{\omega^2}}}  \right].
\end{align}
Note also that the pole at $\gamma = \dw^+$ is not present on the right-hand side of \eq{G2_omega_gamma_constraint0} (and therefore is also not present on the equation's left-hand side). 
Satisfying \eq{G2_omega_gamma_constraint0} would complete the solution of Eqs.~\eqref{largeNc_eqns} by expressing $\gtwg$ in terms of $\gtwg^{(0)}$ and $G^{(0)}_{\omega \gamma}$. Equation~\eqref{G_omegagamma_0} would then allow us to find $G_{\omega\gamma}$, after which all the dipole amplitudes can be constructed using Eqs.~\eqref{G2}, \eqref{Gamma2}, \eqref{G}, \eqref{Gamma}, and \eqref{Gamma_omega_pm}. The quantities $\gtwg^{(0)}$ and $G^{(0)}_{\omega \gamma}$ are specified by the initial conditions/inhomogeneous terms.

The only remaining problem is that \eq{G2_omega_gamma_constraint0} contains $\gtwg$ with two different arguments: it contains $\gtwg$ itself along with $G_{2\omega\dw^+}$. This makes the equation harder to solve for $\gtwg$. However, we can solve \eq{G2_omega_gamma_constraint0} for $G_{2\omega\dw^+}$ by setting $\gamma = \gamma^+_\omega$ in it.  The term $\gtwg\left(\gamma-\gamma^-_\omega\right)\left(\gamma-\gamma^+_\omega\right)$ will vanish as long as $\gtwg$ does not have a pole at $\gamma=\gamma^+_\omega$. However, we assumed this to be true from the beginning: in writing the inverse Laplace transform \eqref{G2}, we assumed that all singularities of $G_{2 \omega \gamma}$ are to the left of the $\gamma$ and $\omega$ integration contours. Since $\gamma^+_\omega \to \omega$ as $\omega \to \infty$, the pole at $\gamma=\gamma^+_\omega$ becomes a pole at $\gamma = \omega$ for large $\omega$. If the $\omega$ and $\gamma$ integration contours were chosen in \eq{G2} such that Re~$\omega >$~Re~$\gamma$, then a pole at $\gamma = \omega$ would violate the assumption of the $\gamma$-contour being to the right of all the singularities of the integrand. If the $\omega$ and $\gamma$ integration contours were chosen such that Re~$\omega <$~Re~$\gamma$, the same argument would apply to the $\omega$ contour. Hence, by writing \eq{G2} we assumed that the pole at $\gamma=\gamma^+_\omega$ in $\gtwg$ does not exist. 

Putting $\gamma = \gamma^+_\omega$ in \eq{G2_omega_gamma_constraint0} and dropping the $\gtwg\left(\gamma-\gamma^-_\omega\right)\left(\gamma-\gamma^+_\omega\right)$  term allows one to solve for $G_{2 \omega \delta_\omega^+}$, yielding
\begin{align}\label{G2_eq_5}
    G_{2 \omega \delta_\omega^+} = \frac{1}{\omega \, \left(\dw^+ - \gamma^-_\omega \right) \, \left(\dw^+ - \gamma^+_\omega \right)} \left\{ 2 \, \left(\dw^+ - \gamma^+_\omega \right) \left[ G_{\dw^+\gamma_\omega^+}^{(0)} + 2G_{2 \, \dw^+\gamma_\omega^+}^{(0)} \right]  - 8 \, \delta_\omega^- \, G_{2 \omega \gamma_\omega^+}^{(0)} + 3 \, \omega \, G_{2 \omega \delta_\omega^+}^{(0)} \right\}.
\end{align}
Substituting this result back into \eq{G2_omega_gamma_constraint0} and solving for $\gtwg$, we obtain
\begin{align}\label{G2_omega_gamma}
    \gtwg = \gtwg^{(0)} + \frac{1}{\omega\left(\gamma-\gamma^-_\omega\right)\left(\gamma-\gamma^+_\omega\right)} &\Bigg[2\left(\gamma-\dw^+\right)\left(G^{(0)}_{\dw^+\gamma} + 2 \, G^{(0)}_{2\dw^+\gamma} \right) \\
    &- 2\left(\gamma^+_\omega-\dw^+\right)\left(G^{(0)}_{\dw^+\gamma^+_\omega} + 2 \, G^{(0)}_{2\dw^+\gamma^+_\omega} \right)
    + 8 \, \dw^-\left(\gtwg^{(0)} - G^{(0)}_{2\omega\gamma^+_\omega} \right) \Bigg]\notag .
\end{align}
Note that indeed, by construction, there is no $\gamma = \gamma^+_\omega$ pole on the right of \eq{G2_omega_gamma}.

We have now completely solved Eqs.~\eqref{largeNc_eqns}. The polarized dipole amplitudes in our solution are given by Eqs.~\eqref{G2}, \eqref{Gamma2}, \eqref{G}, and \eqref{Gamma}, with the ingredients of these expressions constructed in Eqs.~\eqref{delta_omega_pm}, \eqref{Gamma_omega_pm}, \eqref{gamma_omega_pm}, and \eqref{G2_omega_gamma}, for the initial conditions specifying $\gtwg^{(0)}$ and $G^{(0)}_{\omega \gamma}$.


\section{Summary of our Results and the Small-$x$ Asymptotics}

\label{sec:results}

\vspace*{-1mm}
\subsection{Summary of our Results}

Let us now summarize our solution and construct its small-$x$ asymptotics. For brevity, we will utilize the notation defined in \eq{bas_def}. Rescaling
\begin{align}
\omega \to \frac{\omega}{\sqrt{\bas}}, \ \ \ \gamma \to \frac{\gamma}{\sqrt{\bas}}, \ \ \ G_{2\omega\gamma} \to \bas \, G_{2\omega\gamma}, \ \ \  G_{2\omega\gamma}^{(0)} \to \bas \, G_{2\omega\gamma}^{(0)}, \ \ \ G_{\omega\gamma}^{(0)} \to \bas \, G_{\omega\gamma}^{(0)},
\end{align}
with the lowercase $\omega, \gamma$ indices not reflecting the rescaling of those variables, we write our solution as follows:

\begin{mdframed}[linecolor=black!75,backgroundcolor=blue!5, rightmargin=0pt, leftmargin=0pt, linewidth=1.5pt, roundcorner=10pt]
\begin{subequations}\label{alleqs}
\begin{align}
    \label{alleqs_G2}
    &G_2(\xoz^2,zs) = \wint \gint \, e^{\omega\ln(zs\xoz^2) +\gamma \ln \left(\tfrac{1}{\xoz^2\Lambda^2} \right)} \, G_{2\omega\gamma}\,, \\
    \label{alleqs_Gamma2}
    &\Gamma_2(\xoz^2,\xto^2,z's) = \wint\gint \\
    &\qquad\qquad\qquad\qquad \times \left[e^{\omega \ln(z's\xto^2) + \gamma \ln\left(\tfrac{1}{\xoz^2\Lambda^2}\right)}\left(G_{2\omega\gamma} - G^{(0)}_{2\omega\gamma}\right) + e^{\omega\ln(z's\xoz^2) + \gamma \ln\left(\tfrac{1}{\xoz^2\Lambda^2}\right)} \, G^{(0)}_{2\omega\gamma} \right]\,, \notag \\
    \label{alleqs_G}
    &G(\xoz^2,zs) = \wint \gint \, e^{\omega\ln(zs\xoz^2) + \gamma \ln \left(\tfrac{1}{\xoz^2\Lambda^2}\right) } \left[\frac{\omega\gamma}{2 \, \bas}\left(G_{2\omega\gamma} -G^{(0)}_{2\omega\gamma} \right) - 2 \, G_{2\omega\gamma} \right]\,, \\
    \label{alleqs_Gamma}
    &\Gamma(\xoz^2,\xto^2,z's) = \wint \, e^{\omega\ln(z's\xto^2)} \left[\Gamma_{\omega}^+(\xoz^2) \, e^{\dw^+ \ln \left(\tfrac{1}{\xto^2\Lambda^2}\right)} + \Gamma_{\omega}^-(\xoz^2) \, e^{\dw^- \ln \left(\tfrac{1}{\xto^2\Lambda^2}\right)} \right]  \\ 
    & \quad + \wint\gint \notag \, e^{\omega\ln(z's\xto^2) + \gamma \ln \left(\tfrac{1}{\xto^2\Lambda^2}\right)} \left[\frac{\left(-\tfrac{3}{2}\omega\gamma + 4 \, \bas\right)G_{2\omega\gamma} + \tfrac{3}{2}\omega\gamma \, G^{(0)}_{2\omega\gamma}  }{\gamma^2 - \omega\gamma + \bas}  \right]\notag \\
    & \quad -\wint\gint \left[2 \, e^{\omega \ln(z's\xto^2) + \gamma \ln \left( \tfrac{1}{\xoz^2\Lambda^2} \right)}\left(G_{2\omega\gamma} - G^{(0)}_{2\omega\gamma}\right) + 2 \, e^{\omega\ln(z's\xoz^2) + \gamma \ln \left(\tfrac{1}{\xoz^2\Lambda^2} \right)} \, G^{(0)}_{2\omega\gamma}\right],\notag
\end{align}
\end{subequations}
\end{mdframed}

with
\begin{subequations}\label{alleqs2}
\begin{align}
    \label{alleqs_G2_omega_gamma}
    &\gtwg = \gtwg^{(0)} + \frac{\bas}{\omega\left(\gamma-\gamma^-_\omega\right)\left(\gamma-\gamma^+_\omega\right)} \Bigg[2 \left(\gamma-\dw^+\right)\left(G^{(0)}_{\dw^+\gamma} + 2 \, G^{(0)}_{2 \, \dw^+\gamma} \right) \\
    &\hspace{5.5cm}- 2\left(\gamma^+_\omega-\dw^+\right)\left(G^{(0)}_{\dw^+\gamma^+_\omega} + 2 \, G^{(0)}_{2 \, \dw^+\gamma^+_\omega} \right) + 8 \, \dw^-\left(\gtwg^{(0)} - G^{(0)}_{2\omega\gamma^+_\omega} \right) \Bigg], \notag \\
    \label{alleqs_G0}
    &G^{(0)}(\xoz^2,zs) = \wint \gint \, e^{\omega\ln\left(zs\xoz^2\right) + \gamma \ln\left(\tfrac{1}{\xoz^2\Lambda^2}\right)}G^{(0)}_{\omega\gamma}\,, \\
     \label{alleqs_G20}
    &G_2^{(0)}(\xoz^2,zs) = \wint \gint \, e^{\omega\ln\left(zs\xoz^2\right) + \gamma\ln\left(\tfrac{1}{\xoz^2\Lambda^2}\right)}G^{(0)}_{2\omega\gamma}\,, \\
    \label{alleqs_Gamma_plus}
    &\Gamma_{\omega}^+(\xoz^2) = \frac{e^{-\dw^+\ln\left(\tfrac{1}{\xoz^2\Lambda^2}\right)}}{\bas \, (\dw^+-\dw^-)}\gint \, e^{\gamma\ln\left(\tfrac{1}{\xoz^2\Lambda^2}\right)}\frac{\omega \, \dw^+}{2 \, (\gamma-\dw^+)}\\
    &\hspace{5cm}\times\bigg[\gtwg\left(\gamma^2-\omega\gamma + 4 \, \bas- \tfrac{8 \, \bas}{\omega} \, \dw^- \right) - \gtwg^{(0)}\left(\gamma^2-\omega\gamma+4 \, \bas\right)  \bigg], \notag \\
    \label{alleqs_Gamma_minus}
    &\Gamma_{\omega}^-(\xoz^2) = \frac{e^{-\dw^-\ln\left(\tfrac{1}{\xoz^2\Lambda^2}\right)}}{\bas \, (\dw^- -\dw^+)}\gint \, e^{\gamma\ln\left(\tfrac{1}{\xoz^2\Lambda^2}\right)}\frac{\omega \, \dw^-}{2 \, (\gamma-\dw^-)}\\
    &\hspace{5cm}\times\bigg[\gtwg\left(\gamma^2-\omega\gamma + 4 \, \bas-\tfrac{8 \, \bas}{\omega} \, \dw^+  \right) - \gtwg^{(0)}\left(\gamma^2-\omega\gamma+4 \, \bas\right)  \bigg], \notag \\
    \label{alleqs_delta_pm}
    &\dw^\pm = \frac{\omega}{2}\left[1\pm\sqrt{1-\frac{4\,\bas}{\omega^2}} \right], \\
    \label{alleqs_gamma_pm}
    &\gamma^{\pm}_\omega = \frac{\omega}{2}\left[1 \pm \sqrt{1 - \frac{16\,\bas}{\omega^2} \, \sqrt{1-\frac{4\,\bas}{\omega^2}}}  \right].
\end{align}
\end{subequations}

With the four polarized dipole amplitudes known, Eqs.~\eqref{distributions+g1} give us the gluon and (flavor-singlet) quark helicity TMDs and PDFs along with the $g_1$ structure function. We begin by substituting \eq{alleqs_G2} into \eq{gluon_TMD} for the gluon dipole TMD, while neglecting the derivative term at DLA. Integrating out ${\un x}_{10}$ we arrive at
\begin{align}\label{gluon_TMD_3}
     g^{G\,dip}_{1L}(x,k_T^2) = \frac{2 \, N_c}{\as \, \pi^3} \, \frac{1}{k_T^2} \wint \gint \, e^{\omega\ln\left(\tfrac{Q^2}{xk_T^2}\right) + \gamma\ln\left(\frac{k_T^2}{\Lambda^2}\right)} \, 2^{2\omega-2\gamma} \, \frac{\Gamma\left(\omega-\gamma+1\right)}{\Gamma\left(\gamma-\omega\right)} \, \gtwg.
\end{align}
The gluon helicity PDF at DLA follows immediately from substituting \eq{alleqs_G2} into \eq{gluon_PDF},
\begin{align}\label{gluon_PDF_1}
    \Delta G(x,Q^2) = \frac{2 \, N_c}{\as \pi^2} \, \wint \gint \, e^{\omega\ln\left(\tfrac{1}{x}\right) + \gamma\ln\left(\tfrac{Q^2}{\Lambda^2}\right)}\gtwg \,.
\end{align}

Next we substitute Eqs. \eqref{alleqs_G} and \eqref{alleqs_G2} into \eq{quark_TMD} for the flavor-singlet quark helicity TMD (while remembering that $Q=G$ at large $N_c$). Integrating out ${\un x}_{10}$ and $z$ yields\footnote{Note that a special care needs to be taken to extract the DLA part of $g^{S}_{1L}(x,k_T^2)$: this was not done in \eq{quark_TMD_4}.}

\begin{align}\label{quark_TMD_4}
    &g^{S}_{1L}(x,k_T^2) = - \frac{N_f}{\as \, 2 \pi^3} \, \frac{1}{k_T^2}\wint\gint \, \left[  e^{\omega\ln\left(\tfrac{Q^2}{x \, k_T^2}\right) + \gamma\ln\left(\tfrac{k_T^2}{\Lambda^2}\right)} - e^{(\gamma - \omega) \, \ln \left( \tfrac{k_T^2}{\Lambda^2} \right)} \right]\,2^{2\omega-2\gamma}\,\frac{\Gamma\left(1+\omega-\gamma\right)}{\Gamma\left(1-\omega+\gamma\right)} \\ 
    & \hspace*{12cm} \times \,\gamma\left(\gtwg - \gtwg^{(0)}\right)\,. \notag
\end{align}

To obtain the quark helicity PDF in the DLA we substitute Eqs.~\eqref{alleqs_G} and \eqref{alleqs_G2} into \eq{quark_PDF}, again remembering that $Q=G$ at large $N_c$. 
Carrying out the $x_{10}^2$ and $z$-integrals and employing \eq{c3} we arrive at 
\begin{align}\label{quark_PDF_2}
    &\Delta \Sigma(x,Q^2) = - \frac{N_f}{\as \, 2 \pi^2} \, \wint\gint \, \frac{\omega}{\omega-\gamma}\left(\gtwg - \gtwg^{(0)} \right) \, e^{\omega\ln\left(\tfrac{1}{x}\right)} \, e^{\gamma\ln\left(\tfrac{Q^2}{\Lambda^2}\right)} \, ,  
\end{align}
where Re~$\omega >$~Re~$\gamma$ along their contours. 

To obtain the $g_1$ structure function, we replace $N_f \to \tfrac{1}{2} \sum_f Z_f^2$ in \eq{quark_PDF_2}, which gives
\begin{align}\label{g1_3}
    & g_1(x,Q^2)  = - \frac{1}{2} \sum_f Z_f^2 \, \frac{1}{\as \, 2 \pi^2} \, \wint\gint \, \frac{\omega}{\omega-\gamma}\left(\gtwg - \gtwg^{(0)} \right) \, e^{\omega\ln\left(\tfrac{1}{x}\right)} \, e^{\gamma\ln\left(\tfrac{Q^2}{\Lambda^2}\right)} \, ,
\end{align}
again with Re~$\omega >$~Re~$\gamma$ on the integration contours.

Thus in Eqs.~\eqref{gluon_TMD_3}, \eqref{quark_TMD_4}, \eqref{gluon_PDF_1}, \eqref{quark_PDF_2}, and \eqref{g1_3} we have analytic small-$x$ large-$N_c$ expressions for the quark and gluon helicity TMDs, PDFs, and the $g_1$ structure function.


\subsection{Small-$x$ Asymptotics}

Importantly, the small-$x$ asymptotics of the dipole amplitudes in \eq{alleqs} are governed by the rightmost singularity in the complex $\omega$-plane. One can show that this rightmost singularity is a branch point of the large square root in $\gamma^-_{\omega}$. Setting the expression under that large square root in $\gamma^-_{\omega}$ from \eq{alleqs_gamma_pm} to zero gives
\begin{align}\label{intercept_eqn}
    1 - \frac{16\,\bas}{\omega^2} \, \sqrt{1-\frac{4\,\bas}{\omega^2}} = 0\,,
\end{align}
whose rightmost solution in the complex $\omega$-plane is
\begin{align}\label{intercept}
    \omega = \alpha_h \equiv \frac{4}{3^{1/3}} \, \sqrt{\textrm{Re} \left[ \left( - 9 + i \, \sqrt{111} \right)^{1/3} \right] } \,\sqrt{\frac{\as \, N_c}{2 \pi}} \approx 3.66074 \, \sqrt{\frac{\as \, N_c}{2 \pi}}\,.
\end{align}

We arrive at the small-$x$ asymptotics of all the helicity-dependent quantities discussed above, driven by the following leading power of $1/x$:
\begin{align}\label{asympt_all}
    \Delta \Sigma (x, Q^2) \sim \Delta G (x, Q^2) 
    \sim g_1 (x, Q^2) \sim g^{G\,dip}_{1L}(x,k_T^2) \sim g^{S}_{1L}(x,k_T^2) \sim \left( \frac{1}{x} \right)^{\alpha_h}\,.
\end{align}
Together with the general solution of the large-$N_c$ small-$x$ helicity evolution equations given in Eqs.~\eqref{alleqs} and \eqref{alleqs2}, the asymptotics \eqref{asympt_all} are the main result of this work.


\section{Resummed Anomalous Dimension and Cross-Checks}

\label{sec:cross_checks}

Now let us perform several cross-checks of our solution.  As a first cross-check, we consider \cite{Cougoulic:2022gbk} where our Eqs.~\eqref{largeNc_eqns_unscaled} were solved iteratively with the initial conditions $G^{(0)}_2(\xoz^2,zs) = 1 $ and $G^{(0)}(\xoz^2,zs) = 0$. With these initial conditions, Eqs.~\eqref{alleqs_G0}, \eqref{alleqs_G20}, and \eqref{alleqs_G2_omega_gamma} give
\begin{align}\label{specific_ICs}
    G^{(0)}_{\omega\gamma} = 0\,,\hspace{1cm} G^{(0)}_{2\omega\gamma} = \frac{1}{\omega\gamma}\,,\hspace{1cm} \gtwg = \frac{1}{\omega(\gamma-\gamma^{-}_{\omega})}.
\end{align} 
One can then expand Eqs.~\eqref{alleqs} in powers of $\as$ and integrate over $\gamma$ and $\omega$ (it is easier to carry out the $\gamma$-integrals first, then expand in powers of $\as$, then carry out the $\omega$-integrals). We have confirmed up to $\mathcal{O}(\as^2)$ that such an expansion of our analytic solution is in complete agreement with the iterative solution from \cite{Cougoulic:2022gbk}.

As another cross-check we can use the gluon helicity PDF $\Delta G(x,Q^2)$ given in \eq{gluon_PDF_1}. Employing $G^{(0)}_{\omega\gamma} $, $G^{(0)}_{2\omega\gamma}$ and $\gtwg$ from \eq{specific_ICs} in \eq{gluon_PDF_1} we obtain
\begin{align}\label{gluon_PDF_soln2}
    \Delta G(x,Q^2) = \frac{2N_c}{\as \pi^2}\wint e^{\omega\ln\left(\tfrac{1}{x}\right) + \gamma^-_{\omega}\ln\left(\tfrac{Q^2}{\Lambda^2}\right)}\frac{1}{\omega}\,.
\end{align}
We see that $\Delta \gamma_{GG}(\omega) \equiv \gamma^-_{\omega}$ is our prediction for the resummed all-order in $\as$  $GG$ anomalous dimension (at small $x$ and in the large-$N_c$ limit), 
\begin{align}\label{anomalous_dim}
\Delta \gamma_{GG}(\omega) = \gamma^-_{\omega} = \frac{\omega}{2}\left[1 - \sqrt{1 - \frac{16\,\bas}{\omega^2}\sqrt{1-\frac{4\,\bas}{\omega^2}} } \ \right] .
\end{align}
Expanding this in powers of $\as$ we obtain
\begin{align}\label{anomalous_dim_exp}
\Delta \gamma_{GG}(\omega)  = \frac{4\,\bas}{\omega} + \frac{8\,\bas^2}{\omega^3} + \frac{56\,\bas^3}{\omega^5} + \frac{496\,\bas^4}{\omega^7} + \mathcal{O}(\as^5). 
\end{align}
Thus our all-order resummed small-$x$ anomalous dimension $\Delta \gamma_{GG}(\omega)$ agrees with the fixed-order calculations to the existing three-loop order \cite{Altarelli:1977zs,Dokshitzer:1977sg,Mertig:1995ny,Moch:2014sna}, with novel predictions at $\mathcal{O}(\as^4)$ and beyond. This accomplishes another cross-check of our solution.


\section{Comparison to BER}

\label{sec:comp_BER}

Here we compare our results to the earlier resummation for helicity distributions at small $x$ done by Bartels, Ermolaev, and Ryskin (BER) \cite{Bartels:1995iu,Bartels:1996wc}. In order to do so, we will need to simplify the expression for the anomalous dimension obtained by BER in \cite{Bartels:1996wc} for the pure-glue case. Following \cite{Bartels:1996wc} we write the $g_1$ structure function as
\begin{align}
    g_1 (x, Q^2) = - \frac{1}{2 \pi} \, \mbox{Im} \, T_3 (x, Q^2) 
\end{align}
with the signature-odd scattering amplitude $T_3$ given by
\begin{align}\label{T3}
    T_3^S (x, Q^2) = \int \frac{d \omega}{2 \pi i} \, \xi (\omega) \, \left( \frac{1}{x} \right)^\omega \, \left( \frac{Q^2}{\Lambda^2} \right)^{F_0 (\omega) / 8 \pi^2} \, \frac{1}{\omega - F_0 (\omega) / 8 \pi^2} \, R_B
\end{align}
for the flavor-singlet case (denoted by the superscript $S$ on the amplitude). Here
\begin{align}
    \xi (\omega) = \frac{e^{- i \pi \omega} -1}{2} \approx \frac{- i \pi \omega}{2}
\end{align}
is the signature factor, $R_B$ is given by the Mellin transform of the Born initial conditions, while $\Lambda$ is our IR cutoff, denoted by $\mu$ in \cite{Bartels:1996wc}. 

The anomalous dimension $F_0 (\omega) / 8 \pi^2$ was found in \cite{Bartels:1996wc} to be (see Eq.~(4.8) in \cite{Bartels:1996wc})
\begin{align}\label{F0}
    \frac{F_0 (\omega)}{8 \pi^2} = \frac{\omega}{2} \, \left[ 1 - \sqrt{1 - \frac{2 \, \as}{\pi \, \omega^2} \, M_0 + \frac{\as}{\pi^3 \, \omega^3} \, G_0 \, F_8 (\omega)  } \right]. 
\end{align}
Here $M_0$ and $G_0$ are $2\times2$ matrices in the quark-gluon distributions space. Their gluon--gluon components are $(M_0)_{GG} = 4 N_c$ and $(G_0)_{GG} = N_c$. The adjoint (octet) amplitude $F_8 (\omega)$ has to be found by solving the following non-linear differential equation,
\begin{align}\label{F8_equation}
    F_8 (\omega) = \frac{4 \pi \as}{\omega} \, M_8 + \frac{\as \, N_c}{2 \pi \, \omega} \, \frac{d F_8 (\omega)}{d \omega} + \frac{1}{8 \pi^2 \, \omega} \, [F_8 (\omega)]^2 .
\end{align}
Again, $M_8$ is a $2\times2$ matrix in the quark and gluon distributions space: its gluon-gluon component is $(M_8 )_{GG} = 2 N_c$. 

To obtain the pure-glue anomalous dimension we discard quarks, and replace the matrices $M_0$, $G_0$, and $M_8$ by their $GG$ components. The matrix functions $F_0 (\omega)$ and $F_8 (\omega)$ also become single-component objects, which we will label $F_{0 \, GG} (\omega)$ and $F_{8 \, GG} (\omega)$, respectively. The solution of \eq{F8_equation} can then be found by using the substitution \cite{Kirschner:1983di}
\begin{align}
    F_{8 \, GG} (\omega) = 4 \pi \as N_c  \, \frac{\pd}{\pd \omega} \ln u (z)
\end{align}
where 
\begin{align}
    z = \frac{\omega}{\omega_0} \ \ \ \mbox{with} \ \ \ \omega_0 = \sqrt{\frac{\as \, N_c}{2 \pi}}.
\end{align}
This reduces \eq{F8_equation} to
\begin{align}\label{u_eq}
    u''(z) - z \, u'(z) + 2 \, u(z) =0.
\end{align}
The solution of \eq{u_eq} giving the right perturbative expansion of $F_{8 \, GG} (\omega)$ in the powers of $\as$ (that is, giving $F_{8 \, GG} (\omega) = 8 \pi \as \, N_c/\omega$ at order-$\as$) is quite simple, 
\begin{align}
    u(z) = z^2 - 1, 
\end{align}
leading to
\begin{align}
    F_{8 \, GG} (\omega) = \frac{8 \pi \as \, N_c}{\omega} \, \frac{1}{1- \frac{\as \, N_c}{2 \pi} \, \frac{1}{\omega^2}}.
\end{align}
Using this result in \eq{F0} along with $(M_0)_{GG} = 4 N_c$ and $(G_0)_{GG} = N_c$ yields the re-summed $GG$ polarized small-$x$ large-$N_c$ anomalous dimension\footnote{While we did not take the large-$N_c$ limit in our calculation, taking it now would not modify anything in \eq{BER_anom_dim}: it appears that pure glue and large-$N_c$ approximations are identical for BER evolution.}
\begin{align}\label{BER_anom_dim}
    \Delta \gamma^{BER}_{GG} (\omega) \equiv \frac{F_{0 \, GG} (\omega)}{8 \pi^2} = \frac{\omega}{2} \, \left[ 1 - \sqrt{1 - \frac{16 \, \bas}{\omega^2} \, \frac{1 - \frac{3 \, \bas}{\omega^2}}{1 - \frac{\bas}{\omega^2} } } \ \right].
\end{align}
Comparing this with \eq{anomalous_dim}, we conclude that our re-summed $GG$ polarized small-$x$ anomalous dimension is different from the one which follows from the evolution obtained by BER. Curiously, the perturbative expansion of $\Delta \gamma^{BER}_{GG} (\omega)$ in the powers of $\as$ yields
\begin{align}\label{BER_exp}
    \Delta \gamma^{BER}_{GG} (\omega) = \frac{4 \, \bas}{\omega} + \frac{8 \, \bas^2}{\omega^3} + \frac{56 \, \bas^3}{\omega^5} + \frac{504 \, \bas^4}{\omega^7} + {\cal O} (\as^5) .
\end{align}
Comparing this with \eq{anomalous_dim_exp}, we see that the two anomalous dimensions, ours and BER, agree at the one-, two- and three-loop levels, which have also been verified by the perturbative calculations \cite{Altarelli:1977zs,Dokshitzer:1977sg,Mertig:1995ny,Moch:2014sna}. However, at the four-loop level, our and BER anomalous dimensions disagree by a small amount. Disagreement persists at higher orders, reflecting the fact that Eqs.~\eqref{anomalous_dim} and \eqref{BER_anom_dim} contain different functions. 

One may wonder about the agreement between the BER intercept and the one found numerically in \cite{Cougoulic:2022gbk}: both intercepts were reported to be $\alpha_h = 3.66 \, \sqrt{\bas}$ \cite{Bartels:1996wc,Cougoulic:2022gbk}. To find an analytic expression for the intercept in the BER calculation, we need to find the right-most singularity of $\Delta \gamma^{BER}_{GG} (\omega)$. Equating the expression under the square root of \eq{BER_anom_dim} to zero, we see that the rightmost branch point is given by
\begin{align}\label{BER_intercept2}
    \omega = \alpha_h^{BER} \equiv \sqrt{\frac{17 + \sqrt{97}}{2}} \, \sqrt{\frac{\as \, N_c}{2 \pi}} \approx 3.66394 \, \sqrt{\frac{\as \, N_c}{2 \pi}},
\end{align}
as first reported in \cite{Kovchegov:2016zex}.
Comparing this to our \eq{intercept}, we see that the two intercepts are indeed also different, though in both cases the numerical prefactor rounds up to $3.66$. 

We thus observe a difference between the solution of our large-$N_c$ helicity evolution equations and the corresponding results of the BER IREE-based resummation. The difference of the intercept appears to be numerically insignificant. The disagreement between the perturbative expansion of our anomalous dimension in \eq{anomalous_dim_exp} and the expansion in \eq{BER_exp} at the four-loop level implies that potential future perturbative calculations of the $GG$ polarized anomalous dimension at four loops can determine which approach is correct.  

The origin of this apparent disagreement between our calculation and that of BER is not entirely clear. We note here that some questions about the validity of one of the approximations made in \cite{Bartels:1996wc} were raised earlier in Appendix~B of \cite{Kovchegov:2016zex}. The questions addressed the role of non-ladder hard (large transverse momentum) gluons in the IREE for helicity: it appears that in \cite{Bartels:1996wc} BER had stated that such hard gluons cannot contribute in the DLA, while in \cite{Kovchegov:2016zex} a counter-example was constructed for the quark--quark scattering amplitude at the order $\as^3$. Since then, a suggestion has been put forward that such hard-gluon non-ladder contributions can be accounted for in the IREE obtained by BER at the order $\as^3$ by re-defining the ladder to include diagrams with uncut rungs.\footnote{One of the authors (YK) thanks Yoshitaka Hatta and Renaud Boussarie for a very useful discussion on this topic.} Below, in Appendix~\ref{A}, we describe how the order-$\as^3$ non-ladder diagrams from \cite{Kovchegov:2016zex} may yet be included into BER IREE, potentially explaining the agreement between the anomalous dimensions \eqref{anomalous_dim_exp} and \eqref{BER_exp} at the order $\as^3$. However, when trying to apply the same line of reasoning to a diagram at the order $\as^4$ containing hard non-ladder gluons, we run into potential problems and cannot unambiguously incorporate it into BER IREE. This issue at the order $\as^4$ may be a possible explanation of the discrepancy between our \eqref{anomalous_dim_exp} and BER \eqref{BER_exp} anomalous dimensions at four loops. We note once again that a full four-loop calculation of polarized DGLAP anomalous dimensions would unambiguously resolve this discrepancy.


\section{Conclusions}

\label{sec:conc}

To summarize, we note that we have analytically solved the large-$N_c$ equations for small-$x$ helicity evolution derived in \cite{Kovchegov:2015pbl, Cougoulic:2022gbk}. The solution for the polarized dipole amplitudes is given in Eqs.~\eqref{alleqs} and \eqref{alleqs2}. The corresponding hTMDS, hPDFs and the $g_1$ structure function are given by Eqs.~\eqref{gluon_TMD_3}, \eqref{quark_TMD_4}, \eqref{gluon_PDF_1}, \eqref{quark_PDF_2}, and \eqref{g1_3}. Our solution results in the small-$x$ asymptotics \eqref{asympt_all} for helicity TMDs, PDFs, and for the $g_1$ structure function, with the intercept given in \eq{intercept}. 

Remarkably, our large-$N_c$ intercept \eqref{intercept}, while being numerically very close to the one resulting from BER IREE \cite{Bartels:1996wc} given by \eq{BER_intercept2} above, is still different. This difference appears to persist when comparing a numerical solution of the large-$N_c \& N_f$ version of the helicity evolution \cite{Kovchegov:2015pbl, Cougoulic:2022gbk} to the appropriate limit of BER work \cite{AKT}. Moreover, the resummed $GG$ large-$N_c$ small-$x$ polarized anomalous dimensions are different in the two approaches: we obtain
\begin{align}\label{anom_us}
\Delta \gamma_{GG}(\omega) = \frac{\omega}{2}\left[1 - \sqrt{1 - \frac{16\,\bas}{\omega^2}\sqrt{1-\frac{4\,\bas}{\omega^2}} } \ \right]
\end{align}
while the BER IREE formalism gives
\begin{align}\label{anom_BER}
\Delta \gamma^{BER}_{GG} (\omega) = \frac{\omega}{2} \, \left[ 1 - \sqrt{1 - \frac{16 \, \bas}{\omega^2} \, \frac{1 - \frac{3 \, \bas}{\omega^2}}{1 - \frac{\bas}{\omega^2} } } \ \right].
\end{align}
We hope that the future developments in perturbative calculations of the polarized DGLAP anomalous dimensions would result in an expression for $\Delta \gamma_{GG}(\omega)$ at four loops, resolving this discrepancy. In the meantime we note that the less than 1$\%$ difference between the two intercepts and a similarly minor difference in the anomalous dimensions are outside of the precision of phenomenological applications of BER and our formalisms for the foreseeable future.


\section*{Acknowledgments}

\label{sec:acknowledgement}

The authors would like to thank Josh Tawabutr for encouraging discussions. One of the authors (YK) is grateful to Johannes Bluemlein and Sven-Olaf Moch for a very informative correspondence. 

This material is based upon work supported by the U.S. Department of Energy, Office of Science, Office of Nuclear Physics under Award Number DE-SC0004286 and within the framework of the Saturated Glue (SURGE) Topical Theory Collaboration.


 \appendix
 \section{Comparison of some diagrams in BER IREE and in the shock wave approach}
 
 \label{A}
 
 The aim of this Appendix is to speculate on the possible origin of the minor disagreement between the result of BER \cite{Bartels:1996wc} and the solution found here, as manifested in the difference between the intercepts (Eqs.~\eqref{intercept} and \eqref{BER_intercept2}) and the anomalous dimensions (Eqs.~\eqref{anom_us} and \eqref{anom_BER}). Admittedly, the authors of this work are not expert enough in the IREE to make any definitive statements, and our discussion below should be understood as pointing out one potential origin of the discrepancy.  
 
The IREE \cite{Gorshkov:1966ht,Kirschner:1983di,Kirschner:1994rq,Kirschner:1994vc,Bartels:1995iu,Bartels:1996wc,Griffiths:1999dj} are based on evolving in the infrared cutoff on the transverse momenta of the quarks and gluons in a $2 \to 2$ forward scattering amplitude. In the original QCD version \cite{Kirschner:1983di}, the IREE for the Reggeon evolution were based on the following observation: the softest loop momentum integral can be driven either by one or two softest partons in the amplitude. Otherwise the amplitude is not double-logarithmic. (In this Appendix, soft and hard refer to the transverse momentum of the partons.)  If there is one softest parton driving the loop integral, then it must be a gluon, and Gribov's bremsstrahlung theorem \cite{Gribov:1966hs,Gorshkov:1969yy} (also known as the soft-gluon theorem) applies, allowing one to keep only the diagrams where the soft gluon connects to the external legs. Since, by definition, the loop integral involving the bremsstrahlung gluon is the softest, the dependence on the IR cutoff $\Lambda$ enters the expression for the amplitude only through the transverse momentum part of the integral,
\begin{align}\label{IRlog}
\int\limits_{\Lambda^2} \frac{d k_T^2}{k_T^2}.
\end{align}
Differentiating the amplitude with respect to $\ln \Lambda^2$ would remove the contribution in \eq{IRlog}, thus truncating (removing) the soft gluon. 

If the softest loop involves two softest partons, in the Reggeon evolution of \cite{Kirschner:1983di} they must be quarks (to transfer the flavor between the projectile and the target) contributing two opposite ``rails" of the ladder: these softest quarks also contribute the logarithm of the IR cutoff in \eq{IRlog}. Truncating these soft quarks allows one to split the single forward $2 \to 2$ scattering diagram into two sub-diagrams, each of them containing a $2 \to 2$ forward scattering sub-process \cite{Kirschner:1983di}. These two observations led to the IREE for the $2 \to 2$ Reggeon scattering amplitude constructed in \cite{Kirschner:1983di}. Similar logic was applied in \cite{Bartels:1995iu} to construct double-logarithmic evolution equations for the flavor non-singlet helicity-dependent amplitude. The latter evolution was confirmed at large $N_c$ based on the $s$-channel shock wave approach in \cite{Kovchegov:2016zex}.  

In the flavor-singlet helicity evolution case \cite{Bartels:1996wc}, an additional category of diagrams was added: the two soft partons could be gluons, also comprising two opposite ``rails" of the ladder. Hence, for the flavor-singlet helicity evolution in \cite{Bartels:1996wc} one may have one softest parton dominating the softest loop integral in an amplitude, which has to be a gluon, or two partons, which could either be two quarks or two gluons, forming opposite ``rails" of the ladder. A consequence of this statement appears to be that for the evolution in \cite{Bartels:1996wc} to work, there should be no non-ladder hard gluons and no hard-gluon vertex corrections: only soft ``bremsstrahlung" non-ladder gluons are allowed, for which the soft-gluon theorem \cite{Gribov:1966hs,Gorshkov:1969yy} applies. In \cite{Bartels:1996wc}, starting after Eq.~(3.32) and until the end of Sec.~2, an argument is presented which appears to make the case that no such hard non-ladder gluons and vertex corrections exist in the flavor-singlet helicity evolution.

\begin{figure}[ht]
\begin{center}
\includegraphics[width= \textwidth]{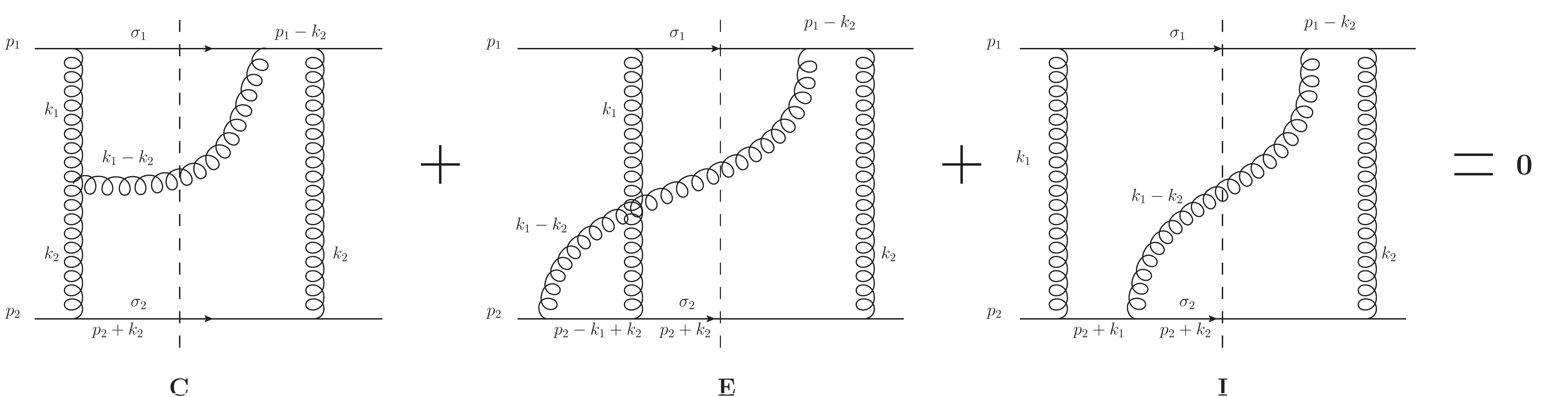} 
\caption{Diagram cancellations in the $k_{2 T} \gg k_{1 T}$ regime with the DLA accuracy, as outlined in \cite{Kirschner:1983di}. We use the diagram labeling from Appendix~B of \cite{Kovchegov:2016zex}. The diagrams are different by different connections of the soft gluon $k_1$ in the lower left corner. Vertical dashed line denotes the cut.}
\label{Feyn_gauge_canc1}
\end{center}
\end{figure}

To better understand this ``no hard non-ladder gluons" assertion, in Appendix~B of \cite{Kovchegov:2016zex} the types of non-ladder gluons were studied by an explicit calculation of several diagrams contributing to the helicity-dependent part of $qq \to qq$ forward scattering at the order $\as^3$. One such diagram, diagram~C in the nomenclature of \cite{Kovchegov:2016zex}, is shown here, in the left panel of \fig{Feyn_gauge_canc1}. The diagram is double-logarithmic: in the $p_1^+, p_2^- \gg k_1^+, k_2^-, k_{1 T}, k_{2 T} \gg k_1^-, k_2^+$ regime it gives \cite{Kovchegov:2016zex} (with the center-of-mass energy squared $s \approx 2 p_1^+ p_2^-$)
\begin{align}\label{Ceq}
\int\limits_{\Lambda^2}^s d k_{1 T}^2  \int\limits_{\Lambda^2}^s \, d k_{2 T}^2 \ C \sim \int\limits_{\Lambda^2}^s \frac{d k_{1 T}^2}{k_{1 T}^2}  \int\limits_{\Lambda^2}^s \, \frac{d k_{2 T}^2}{k_{2 T}^2}
\end{align}
along with a logarithm of energy resulting from the longitudinal momentum integration. (For simplicity we imagine working in a frame with $p_{1 T} = p_{2 T} =0$. The contribution C in \cite{Kovchegov:2016zex} does not explicitly include any of the integrals.)

It is our understanding that in IREE the diagram C from the left panel of \fig{Feyn_gauge_canc1} should be separately considered in two different kinematic regions, $k_{1 T} \ll k_{2 T}$ and $k_{1 T} \gg k_{2 T}$. In either kinematic region, the non-ladder gluon $k_1 - k_2$ is hard, thus contributing a hard (cut) vertex correction in an apparent violation of the absence of such gluons in the IREE argued in \cite{Bartels:1996wc}. However, before reaching any conclusions, let us analyze this diagram C in more detail. 

In the $k_{1 T} \ll k_{2 T}$ region the $k_1$ gluon is the softest in the diagram and the bremsstrahlung theorem applies. Following Sec.~3.3 of \cite{Kirschner:1983di}, we see that diagram~C for $k_{1 T} \ll k_{2 T}$ falls under the category of Fig.~7 in that reference, with the cut through the quark ($p_2 + k_2$) and gluon ($k_1 - k_2$) lines connecting an external leg to the rest of the diagram and with the soft uncut gluon ($k_1$) attaching in all possible ways to the three lines involved ($p_2$, $p_2 + k_2$, and $k_1 - k_2$). (The cut is mentioned in the text, but not shown explicitly in Fig.~7 of \cite{Kirschner:1983di}.) These connections of the soft gluon $k_1$ are shown here in the diagrams C, E and I in \fig{Feyn_gauge_canc1}, using the diagram labeling from \cite{Kovchegov:2016zex}. Employing Eqs.~(B2) from \cite{Kovchegov:2016zex} we readily obtain (division by 4 and 2 is required to single out one diagram in the class of diagrams C, E and I, with the diagrams in each class related to each other by up-down and left-right symmetries)
\begin{align}\label{cancel1}
\left[ \frac{C}{4} + \frac{E}{4} + \frac{I}{2} \right]_{k_{2 T} \gg k_{1 T}} = g^6 \, C_F \, \sigma_1 \, \sigma_2 \, \frac{s}{k_{1 T}^2 \, k_{2 T}^2} \, \left[ -2 + 2 \, \frac{N_c^2 -2}{N_c^2} + \frac{4}{N_c^2} \right] =0. 
\end{align}
Here, as in \cite{Kovchegov:2016zex}, we keep only the part of the amplitude dependent on the polarizations $\sigma_1$ and $\sigma_2$ of the two colliding quarks. Also, $C_F = (N_c^2 -1)/(2 N_c)$ is the fundamental Casimir operator and $g$ is the QCD coupling.

\begin{figure}[ht]
\begin{center}
\includegraphics[width= 0.75 \textwidth]{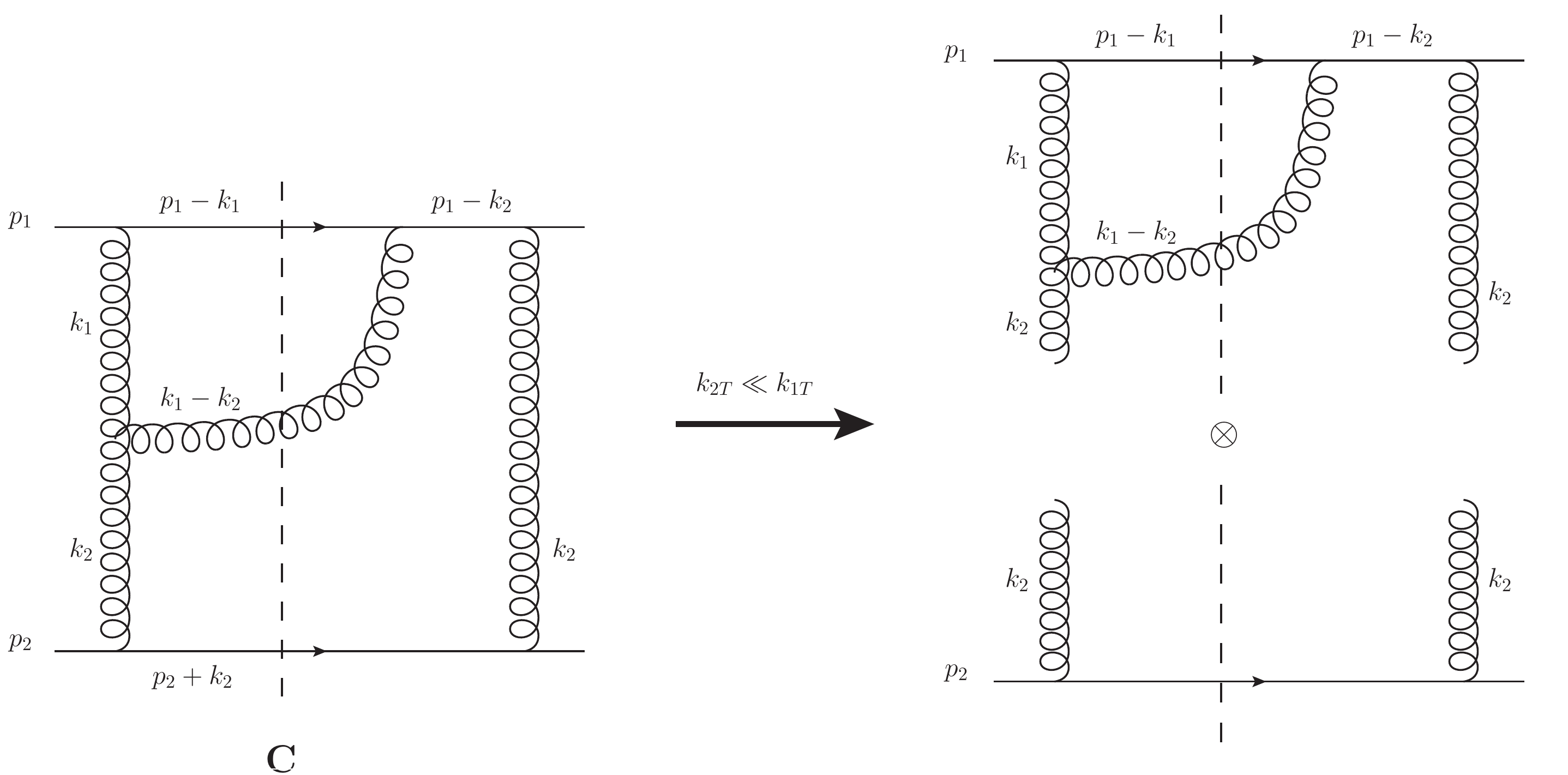} 
\caption{Possible decomposition of the diagram C from Appendix~B of \cite{Kovchegov:2016zex} in the $k_{2 T}  \ll k_{1 T}$ regime under the IREE from \cite{Bartels:1996wc}. Vertical dashed line denotes the cut.}
\label{fig:diagramC}
\end{center}
\end{figure}



In the $k_{1 T} \gg k_{2 T}$ region, both $k_2$ gluons become the softest internal partons in the diagram. According to IREE, one should then truncate these two gluons splitting diagram C into two, as shown in \fig{fig:diagramC}. The diagram on the top right of that figure, resulting from this splitting, appears to still be a non-ladder diagram with a hard gluon $k_1 - k_2$. It still appears to violate the IREE rules. However, to better understand IREE, imagine attaching a bremsstrahlung gluon to a single-rung ladder (an ``H"-shape diagram). By IREE rules, the transverse momentum of the external legs is negligibly small. Therefore, the transverse momentum in the loop formed by attaching a bremsstrahlung gluon to the single-rung ladder is going to be larger than that in the external legs, and, by momentum conservation, should be the same in the propagators of all the partons forming the loop. For the top right diagram in \fig{fig:diagramC} this implies that, in the $k_{1 T} \gg k_{2 T}$ regime, the transverse momenta of the gluon lines $k_1$ and $k_1 - k_2$ are comparable to each other and to the transverse momentum in the quark line $p_1 - k_1$. Therefore, at this low order in $\as$ (order-$\as^2$), there appears to be no difference between a diagram with a bremsstrahlung gluon and a diagram with a hard non-ladder gluon: the top right diagram in \fig{fig:diagramC} can be viewed as a bremsstrahlung gluon diagram. The remaining question is to identify which gluon is the bremsstrahlung one in the top right diagram of \fig{fig:diagramC}: is it the gluon $k_1$ or $k_1 - k_2$? According to \cite{Kirschner:1983di}, the bremsstrahlung gluon should carry longitudinal (``nonsense") polarization. Since the gluon $k_1 - k_2$ is cut, it can only be polarized transversely, and, hence, cannot be the bremsstrahlung gluon. This leaves $k_1$ to be the bremsstrahlung gluon. Therefore, we can view top right diagram in \fig{fig:diagramC} as the ladder made out of the gluons $k_1 - k_2$ and $k_2$ and the quark line, with the rung of the ladder given by the $p_1 - k_2$ quark line, and with the $k_1$ bremsstrahlung soft gluon attached to the ladder. (It is not clear to the authors whether such a ladder with an uncut rung was intentionally included in the BER formalism.) Therefore, while initially appearing to violate the ``no hard non-ladder gluons"  argument, diagram C (along with the diagram B from Appendix~B of \cite{Kovchegov:2016zex}) can be incorporated into the IREE derived by BER. Identifying the forward $2 \to 2$ quark and gluon scattering amplitudes with the anomalous dimensions at the same order in $\as$, per \cite{Kirschner:1983di}, we see that the above discussion appears to explain why the calculation of \cite{Bartels:1996wc} agrees with the polarized DGLAP anomalous dimensions to three loops \cite{Blumlein:1996hb}, that is, to order $\as^3$. 

\begin{figure}[ht]
\begin{center}
\includegraphics[width= \textwidth]{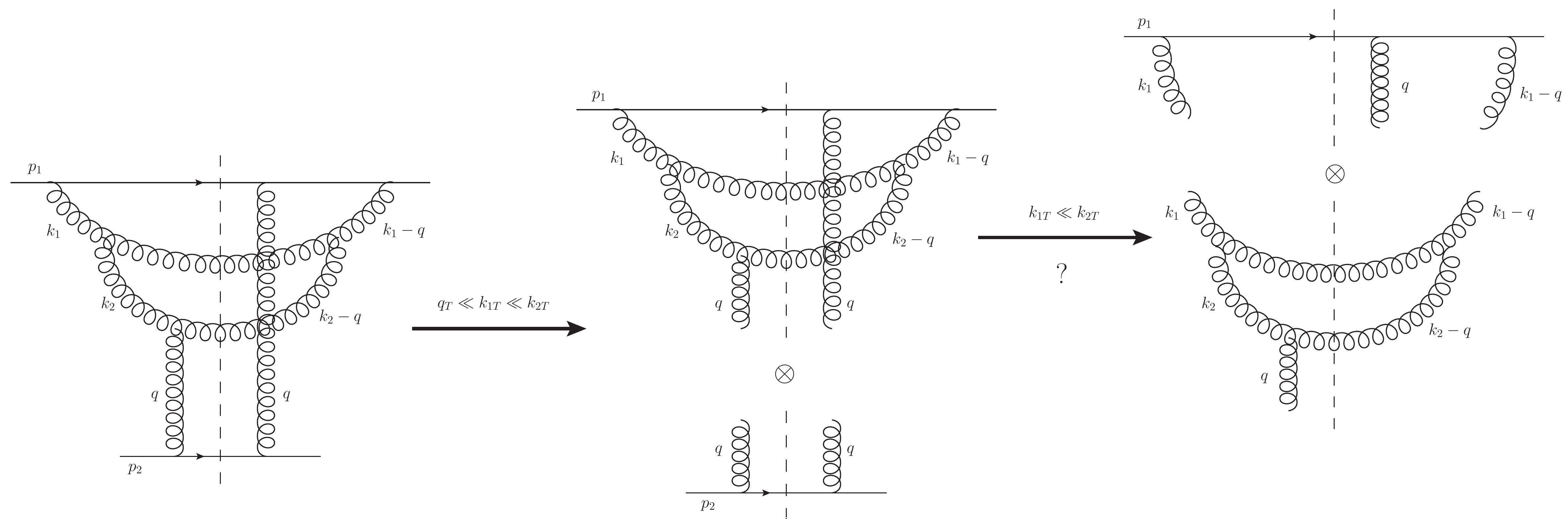} 
\caption{Possible decomposition of a diagram at the order $\as^4$ in the $q_T \ll k_{1 T} \ll k_{2 T}$ regime under the IREE from \cite{Bartels:1996wc}.}
\label{fig:4loops}
\end{center}
\end{figure}

We now want to investigate the hard non-ladder diagrams at higher orders in $\as$. Anomalous dimensions at four loops correspond to $2 \to 2$ forward scattering amplitudes at the order $\as^4$. While a systematic analysis of all order-$\as^4$ diagrams appears to be rather lengthy, we will consider one relevant diagram to illustrate a possible concern arising at that order. This order-$\as^4$ diagram is given in the left panel of \fig{fig:4loops}. This diagrams is known to be double-logarithmic, at least in the framework of \cite{Kovchegov:2015pbl, Kovchegov:2016zex, Kovchegov:2017lsr, Kovchegov:2018znm,Cougoulic:2022gbk}. Note that in \cite{Kovchegov:2015pbl, Kovchegov:2016zex, Kovchegov:2017lsr, Kovchegov:2018znm,Cougoulic:2022gbk} the calculations were performed in the light-cone gauge of the projectile (the upper quark line in \fig{fig:4loops}): it is possible that the diagram in \fig{fig:4loops} is not double-logarithmic in Feynman gauge employed in \cite{Bartels:1996wc}. However, the IREE technique developed in \cite{Kirschner:1983di} is stated to be gauge-invariant by the authors. Hence we proceed by assuming that the diagram in \fig{fig:4loops} is double-logarithmic in Feynman gauge as well. (The authors of \cite{Cougoulic:2022gbk} have also verified their calculations in background Feynman gauge.)


We will concentrate on the kinematic region where $q_T \ll k_{1 T} \ll k_{2 T}$. In this region the diagram in the left panel of \fig{fig:4loops} is still double-logarithmic \cite{Cougoulic:2022gbk}. The diagram appears to be a one-loop vertex correction to the diagram B from Appendix~B of \cite{Kovchegov:2016zex}, which, in turn, is quite similar to the diagram C considered above. In the $q_T \ll k_{1 T} \ll k_{2 T}$ kinematic region, there are two softest gluons, the ones carrying momenta $q$ in \fig{fig:4loops}. (These are the Glauber gluons in the formalism of \cite{Kovchegov:2015pbl, Kovchegov:2016zex, Kovchegov:2017lsr, Kovchegov:2018znm,Cougoulic:2022gbk}.) Truncating those gluons splits the left diagram into the two diagrams in the middle panel of \fig{fig:4loops}. Just like in \fig{fig:diagramC}, the resulting top diagram appears to be non-ladder. One may wonder whether it is also included in the BER evolution. This appears to be less clear. In the same $q_T \ll k_{1 T} \ll k_{2 T}$ kinematic region, the softest two gluons in the top diagram in the middle panel of \fig{fig:4loops} are $k_1$ and $k_1 -q$. The next question is whether these gluons are (i) ``rails" of some ladder or whether (ii) one of them is a bremsstrahlung gluon. In the case (i), by the IREE rules, truncating those gluons leads to the diagrams on the right of \fig{fig:4loops}. However, these diagrams form 3- and 5-point Green functions. The IREE of \cite{Bartels:1996wc} only contain 4-point Green functions and do not contain diagrams with an odd number of external legs. Therefore, if the ladder ``rails" interpretation from (i) is correct, it appears impossible to obtain the contribution of the diagram on the left of \fig{fig:4loops} in the $q_T \ll k_{1 T} \ll k_{2 T}$ kinematic region using IREE. 

The option (ii), involving a bremsstrahlung gluon, appears to be in-line with our above interpretation of the top right diagram in \fig{fig:diagramC}. However, the top middle diagram in \fig{fig:4loops} has a significant difference from the top right diagram in \fig{fig:diagramC}: both gluons $k_1$ and $k_1 -q$ are not cut. Either of them may carry longitudinal polarization. Therefore, it appears unclear, at least to the authors, which of these two gluons would be the bremsstrahlung one. Moreover, the position of gluons $k_1$ and $k_1 -q$ in the diagram appears to be rather ladder-like, making applicability of Gribov's bremsstrahlung theorem \cite{Gribov:1966hs,Gorshkov:1969yy} questionable; after all, the bremsstrahlung theorem does not apply to two equally soft gluons forming the ``rails" of a ladder. It, therefore, appears unlikely that the diagram on the left of \fig{fig:4loops} in the $q_T \ll k_{1 T} \ll k_{2 T}$ kinematic region can be obtained from the IREE developed in \cite{Bartels:1996wc}.

However, our admittedly limited understanding of IREE does not allow us to reach a firm conclusion here. Indeed it is also possible, though perhaps unlikely, that the diagram on the left of  \fig{fig:4loops} is not double-logarithmic in Feynman gauge. Alternatively, it may also be possible to interpret this diagram in BER IREE using some observation currently not apparent to the authors of this work. Yet again, in \cite{Bartels:1996wc}, BER do express concern about hard non-ladder diagrams and appear to argue that those are not double-logarithmic: the apparent violation of that argument found in \cite{Kovchegov:2016zex} should manifest itself at some order in $\as$. Moreover, if our concern expressed in this Appendix is correct, the fact that it applies only to a fairly high-order in $\as$ diagram may explain the numerically minor difference of the intercepts in Eqs.~\eqref{intercept} and \eqref{BER_intercept2} and the fact that the expansions \eqref{anomalous_dim_exp} and \eqref{BER_exp} for our and BER anomalous dimensions disagree only starting at the order $\as^4$.



\providecommand{\href}[2]{#2}\begingroup\raggedright\endgroup

\end{document}